\documentclass{aa}  

\usepackage{graphicx}
\usepackage{txfonts}
\usepackage{lipsum}
\usepackage{subcaption}         
\usepackage{lscape}             
\usepackage{placeins}           
\usepackage{xcolor}
\usepackage{diagbox}

\usepackage[breaklinks=true]{hyperref} 

\begin{document}

   \title{Star--planet interaction in the Proxima system}


   \author{M$.$ R$.$ Zapatero Osorio\inst{1}\thanks{e-mail: mosorio@cab.inta-csic.es}
        \and V$.$ J$.$ S$.$ B\'ejar\inst{2,3}
        \and A$.$ Su\'arez Mascare\~no\inst{2,3}
        \and R$.$ Rebolo\inst{2,3,4}
        \and S$.$ Cristiani\inst{5,6,7}
        \and G$.$ Micela\inst{8}
        \and P$.$ A$.$ Miles-P\'aez\inst{1}
        \and N$.$ C$.$ Santos\inst{9,10}
        \and E$.$ Pall\'e\inst{2,3}
        \and J$.$ I$.$ Gonz\'alez Hern\'andez\inst{2,3}
        \and M$.$ Damasso\inst{11}
        \and A$.$ Castro-Gonz\'alez\inst{12}
        \and C$.$ J$.$ A$.$ P$.$ Martins\inst{9,13}
        \and F$.$ Pepe\inst{12}
        \and A$.$ Sozzetti\inst{11}
        \and B$.$ Lavie\inst{12}
        \and J$.$ Rodrigues\inst{9,10,12}
        }

   \institute{Centro de Astrobiolog\'\i a, CSIC-INTA, Camino Bajo del Castillo s/n, 28692 Villanueva de la Ca\~nada, Madrid, Spain\\ 
             \and Instituto de Astrof\'\i sica de Canarias, C/ V\'\i a L\'actea s/n, 38205 La Laguna, Tenerife, Spain 
             \and Departamento de Astrof\'\i sica, Universidad de La Laguna, 38206 La Laguna, Tenerife, Spain 
             \and Consejo Superior de Investigaciones Cient\'\i ficas (CSIC), 28006 Madrid, Spain 
             \and INAF -- Osservatorio Astronomico di Trieste, via G.B. Tiepolo, 11, I-34143 Trieste, Italy 
             \and IFPU -- Institute for Fundamental Physics of the Universe, via Beirut 2, I-34151 Trieste, Italy 
             \and INFN -- National Institute for Nuclear Physics, via Valerio 2, I-34127 Trieste, Italy 
             \and INAF -- Osservatorio Astronomico di Palermo, 1 Piazza del Parlamento, Palermo 90134, Italy 
             \and Instituto de Astrof\'{\i}sica e Ci\^encias do Espa\c co, Universidade do Porto, Rua das Estrelas, 4150-762 Porto, Portugal 
             \and Dept$.$ de F\'\i sica e Astronomia, Faculdade de Ciencias, Universidade do Porto, Rua Campo Alegre, 4169-007 Porto, Portugal 
             \and INAF -- Osservatorio Astrofisico di Torino, Via Osservatorio 20, 10025 Pino Torinese, Italy 
             \and Observatoire de Gen\`eve, D\'epartement d’Astronomie, Universit\' e de Gen\`eve, Chemin Pegasi 51b, 1290 Versoix, Switzerland 
             \and Centro de Astrof\'{\i}sica da Universidade do Porto, Rua das Estrelas, 4150-762 Porto, Portugal 
             }
   \date{Received }

  \abstract
   {}
   {We search for evidence of star--planet magnetic interactions in the nearby Proxima Centauri planetary system using high-quality, high-spectral-resolution optical observations spanning three years.} 
   {We investigate the variability of photospheric and chromospheric features traced by oxides, H\,{\sc i}, Na\,{\sc i}, Fe\,{\sc i}, He\,{\sc i}, and Ca\,{\sc ii}  using a differential spectroscopic method and narrow-band light curves derived from the spectra.}
   {We measure a photospheric stellar rotation period of 84.9\,$\pm$\,0.6 d and a half-rotation period of 44.3\,$\pm$\,0.2 d, consistent with previous studies. The stellar photospheric absorption lines show no variability correlated with the orbital periods of Proxima b or d, thus supporting their planetary nature. Using Fe\,{\sc i} absorption and emission lines, we find that Proxima Centauri was flaring during 24.8 $\pm$ 4.7\,\%~of the observing time, with significant statistical evidence ($\ge$99.8\,\%) of flare events likely phase-locked to the inner Mars-mass planet Proxima d. Modeling the star--planet interaction via the helicity-driven reconnection mechanism with the Poynting flux formalism, we estimate a likely polar magnetic field of  $\sim$16 G for Proxima d (assuming a Mars-sized radius), with a plausible range of 3--280 G accounting for radial and dipolar stellar magnetic field configurations, planetary radii comparable to Mars and Earth, and the observed range of stellar flare intensities. This represents the first such estimate for a terrestrial exoplanet. Evidence for a potential star--planet interaction with the outer, Earth-mass Proxima b arises not from phase-locked flare clustering, but from modulation of flare intensities. Applying a prewhitening analysis to the full time series of combined chromospheric H$\alpha$, Na\,{\sc i} D1 \& D2, and Ca\,{\sc ii} H \& K lines reveals peaks, in order, at half the stellar rotation period, Proxima b’s orbital period, the full stellar rotation, and Proxima d’s orbital period. All evidence suggests that both planets show magnetic interaction with their host star. Focusing on flaring epochs only, the periodogram of these chromospheric lines shows a peak consistent with the synodic period between half the stellar rotation and the mutual synodic period of Proxima b and d, implying prograde stellar rotation and planetary orbits. We also identify spectroscopic features at around 7794.0 and 7808.5 \AA~that behave differently from the rest of the spectrum.
}
   {}


   \maketitle

\nolinenumbers

\section{Introduction}

Proxima Centauri is the nearest star to the Sun (1.30197 $\pm$ 0.00008 pc; \citealt{gaia23}) and hosts at least two close-in terrestrial planets, Proxima Centauri b with a minimum mass of $M_{\rm p}$\,sin\,$i$ = 1.3 M$_\oplus$ and Proxima Centauri d with $M_{\rm p}$\,sin\,$i$ = 0.26 M$_{\oplus}$ (Proxima b and Proxima d from now on), with orbital periods of 11.1 d and 5.1 d, respectively \citep{escude16, faria22, mascareno25}. A third, more distant planet candidate, Proxima Centauri c, was reported by \citet{damasso20a} but remains unconfirmed. Proxima Centauri is an M5.5V main-sequence star, a member of a hierarchical triple system \citep{kervella17}, with approximately solar metallicity and an estimated age similar or slightly older than the Sun \citep{mamajek08, ribas16, ribas17, thevenin26}. Since the discovery of the 11.1-d Proxima b, located within the stellar habitable zone \citep{escude16}, the system has been extensively characterized across the electromagnetic spectrum. At 1.3 mm, \citet{anglada17} reported a cold dust belt at $\sim$1--4 au, analogous to the Solar System’s Kuiper Belt. Using high-resolution optical and near-infrared spectroscopy, \citet{silva25} have probed atmospheric diagnostics sensitive to stellar variability, while precise radial velocity monitoring led to the detection of the 5.1-d Proxima d \citep{mascareno20, faria22}. In the ultraviolet and X-rays, \citet{sanzforcada25} compiled archival observations to reconstruct the stellar coronal emission, while \citet{damonte26} produced stellar spectra from 1 to 920 \AA, revealing wavelength-dependent X-ray variability.

Proxima Centauri is magnetically active and a well-known flare star \citep{shapley51}, producing frequent eruptions over a wide range of energies from the far ultraviolet to the millimeter wavelengths  \citep[e.g.,][]{loyd18, howard18, macgregor21, burton25}. According to \citet{pavlenko19}, all optical emission lines show variability with a timescale of at least 10 min. Using space-based observations from the {\sl Transiting Exoplanet Survey Satellite} ({\sl TESS}; \citealt{ricker15}) and the {\sl Microvariability and Oscillations of STars} microsatellite \citep{walker03}, together with extensive ground-based photometric monitoring campaigns, various studies have quantified the cumulative flare frequency of Proxima Centauri in the optical band \citep[e.g.,][]{davenport16, vida19, macgregor21}. For flare energies between $\sim$10$^{+30.5}$ and $\sim$10$^{+33.5}$ erg, these works found an overall occurrence rate of approximately 1.5--5 flares per day when considering all energies and amplitudes. In contrast, for flare energies spanning $\sim$10$^{+24}$ to $\sim$10$^{+27}$ erg in the millimeter regime, the cumulative flare frequency shows a much steeper slope, corresponding to nearly 300 low-energy, short-duration flares per day \citep{burton25} that may play a role in coronal heating on Proxima Centauri. The analysis of {\sl TESS} optical data by \cite{vida19} further showed that only two out of 72 detected flares had durations exceeding several hours, while the vast majority lasted for less than a few hours. This contrasts with the shorter duration of the millimeter flares, which range from seconds to minutes \citep{burton25}. These studies attributed the flares to intrinsic stellar activity and reported no evidence for periodicity in the flare occurrence. The same {\sl TESS} data were used by \citet{vida19} and \citet{gilbert21} to confirm the non-transiting nature of Proxima b and d. Characterizing flare activity is fundamental for understanding atmospheric erosion and retention on orbiting planets, and  for assessments of habitability around M-type dwarf stars (e.g., \citealt{engelbrecht24, pena24, zhu25}). Proxima Centauri serves as a benchmark system for studying the atmospheres of terrestrial planets in M-dwarf environments because of its proximity

The system has also been investigated in the context of star--planet interactions, although a controversy persists regarding whether Proxima b resides in the sub- or super-Alfv\'enic regime \citep[e.g.,][]{klein21, pena25}. A confirmed detection of star--planet interaction signatures in the Proxima Centauri system remains elusive despite extensive observational and theoretical efforts. \citet{pereztorres21} reported circularly polarized radio emission modulated at approximately half the orbital period of Proxima b; these authors interpreted this signal as evidence that the planet orbits in the sub-Alfv\'enic regime. This interpretation is consistent with the modeling of \citet{garraffo22} and \citet{reville24} and with predictions that coherent radio emission should arise from sub-Alfv\'enic interactions \citep{turnpenney18}, but is at odds with the findings of \citet{kavanagh21}. At optical wavelengths, the star--planet connection has been addressed through flare events observed with {\sl TESS} by \citet{vida19} and \citet{ilin24}, yielding no conclusive evidence. Very recently, \citet{ilin25} reported the first detection of planet-induced flares on HIP\,67522, a 17 Myr old G dwarf star with two known close planets. There have also been claims of X-ray flare-intensity modulation linked to the presence of a giant planet around HD\,189733 \citep[][and references therein]{pillitteri22}.

Here, we report an analysis of photospheric and chromospheric atomic and molecular features from high-resolution archival optical spectra, revealing likely evidence of star--planet interactions for both close-in Proxima planets.

\section{Observations}

We used 117 high-quality, high spectral resolution observations ($R$\,$\sim$\,134,000) obtained with the Echelle SPectrograph for Rocky Exoplanets and Stable Spectroscopic Observations (ESPRESSO; \citealt{pepe21}) on the Very Large Telescope at Cerro Paranal, Chile. These data, previously presented by \citet{mascareno20} and \citet{faria22}, were collected over nearly three years, with a 1.7 year gap due to the COVID-19 pandemic and the subsequent restart of ESPRESSO operations. We analyzed the rebinned and merged one-dimensional spectra (3782.5--7886.5 \AA), corrected for instrumental response. The ESPRESSO spectra were also corrected for telluric absorption from H$_2$O and O$_2$ using the publicly available \texttt{TelFit} code \citep{gullikson14}. We developed a custom routine to iteratively determine the optimal molecular abundances for each epoch. The routine removes telluric lines by minimizing the root mean square (rms) of residuals (observation minus telluric model) within selected wavelength regions affected by Earth’s atmospheric absorption. However, the optical spectra of mid-M dwarfs are dominated by strong TiO and VO molecular absorption features originating in their cool atmospheres, which are often blended with telluric lines. This blending generally leads to an underestimation of the true atmospheric water and oxygen abundances, and residual telluric absorption features may therefore remain in the corrected spectra. Note that the routine does not remove telluric emission lines. However, these lines are few and can be easily identified because they shift with the barycentric velocity.

\section{Analysis}
Our analysis is based on the differential spectroscopy technique. We first constructed a high-quality reference spectrum by median-combining all 117 single-epoch observations. Differential spectra were then obtained by subtracting each individual spectrum from this reference spectrum. The procedure consists of the following steps:
\begin{itemize}
\item To minimize large-scale effects caused by the continuum wavelength-dependent absorption of the Earth’s atmosphere (since the data were acquired at different airmasses and under varying weather conditions), we divided each spectrum into chunks ranging from a few to several hundred angstroms in width. The chunk size is generally smaller at blue wavelengths, where differential extinction has a stronger impact. Chunk boundaries were chosen to avoid overlap with known chromospheric lines.
\item Each chunk was then normalized by its median flux after excluding flux values that deviated by more than $1.4\,\sigma$ above and $1.8\,\sigma$ below the median. This step prevents strong emission and absorption features from biasing the normalization.
\item The normalized spectra were subsequently resampled onto a common wavelength grid and median-combined. The uncertainties of the combined spectral fluxes were computed as the median of the individual photon-noise errors.
\item Each individual normalized spectrum was then subtracted from the reference spectrum to create the differential data used throughout the paper. The error bars for each differential flux were calculated as the quadratic sum of the uncertainties from both the original and reference spectra.
\item Differential spectra are affected by interference patterns that appear as sinusoidal “wiggles” across all wavelengths \citep{allart20, borsa21, tabernero21, casasayas21}. "Wiggles" arise from the coud\'e train optics and vary with the telescope pointing. The literature reports two types: short-period "wiggles" with a $\sim$1 \AA\ period and $\sim$0.1\,\%~amplitude at 600 nm, and long-period "wiggles" with 30--40 \AA\ periods and amplitudes approximately ten times larger. We did not correct for the short-period "wiggles", but we applied an iterative sigma-clipped spline fit to correct the large-amplitude component in all differential spectra. In each iteration, a median filter estimates the continuum, and fluxes deviating by more than 4 $\sigma$ are rejected. A spline is then fitted to the remaining points, and the continuum is subtracted. This process is repeated until convergence or until a maximum number of iterations is reached. An example of the "wiggles" correction is shown in Figure~\ref{fig:wiggles}. This step flattened all differential spectra, effectively removing the long-period "wiggles" while preserving narrow-band variability. Note that the process also removed large-scale wavelength variations intrinsic to Proxima Centauri, but this does not impact the objectives of this work.
\item Finally, we focused on specific spectral features and extracted their light curves by summing the fluxes within defined wavelength ranges. Since all differential spectra are relative to the reference spectrum, positive and negative flux sums correspond to increased emission and absorption, respectively. The wavelength width varies for each feature, ranging from approximately 0.05 \AA\ to 0.2 \AA. The uncertainty for each flux sum is based on the mean errors of the wavelength interval.
\end{itemize}

After this procedure, we typically reached flux dispersions of $\pm510$ parts per thousand (ppt) at $\sim$4000~\AA, $\pm 35$ ppt at $\sim$5450~\AA, and $\pm 6.1$ ppt at 7800~\AA~for the 117 differential spectra altogether. The precision improves toward red wavelengths, where Proxima Centauri is brighter.

\begin{figure}[h!]
\centering
\includegraphics[width=0.99\hsize]{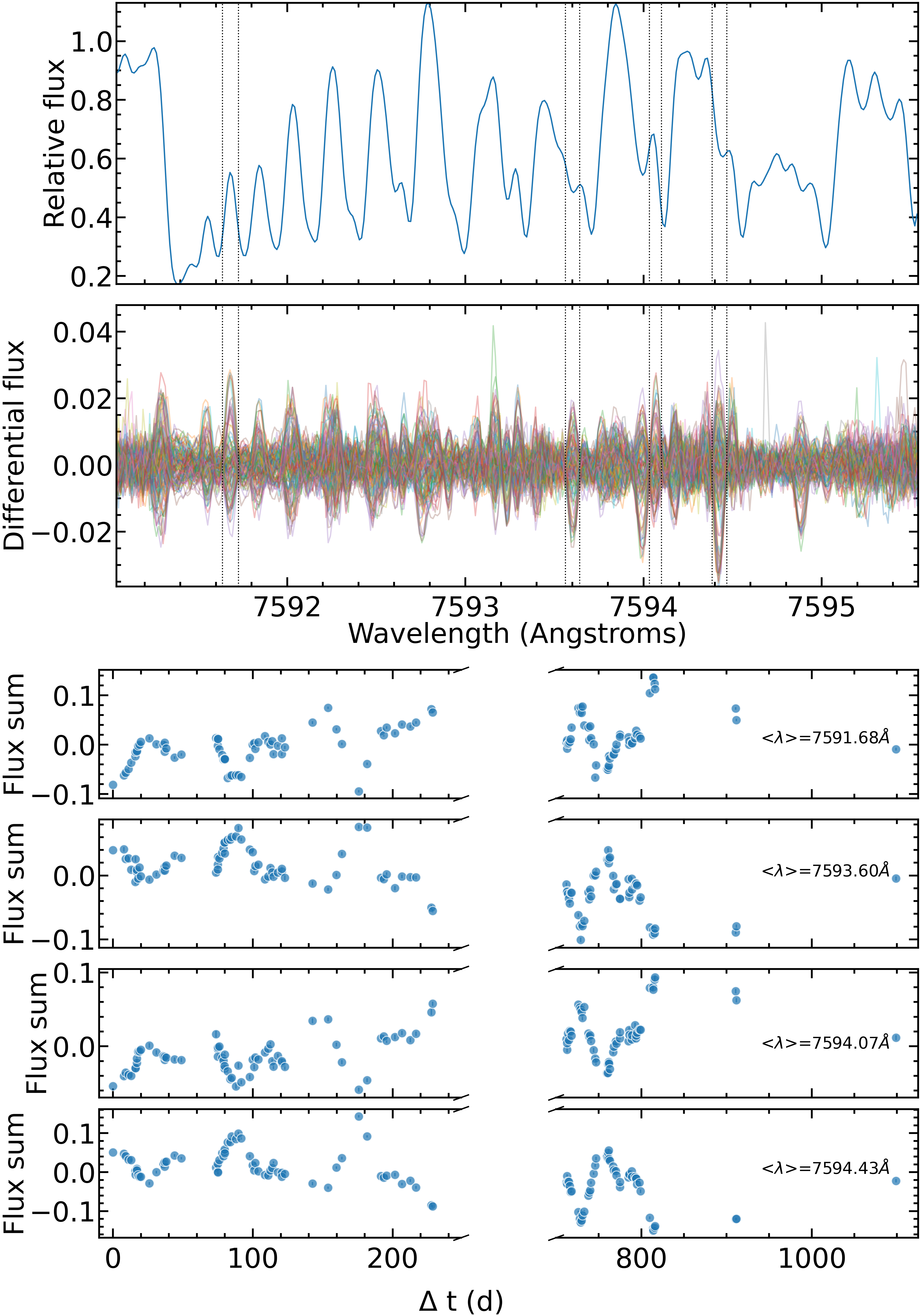}
\caption{A small portion of the combined ESPRESSO (reference) spectrum of Proxima Centauri is shown in the first panel. Most of the spectral features are due to TiO absorption. The second panel displays the 117 individual differential spectra. Notable spectral variability due to stellar rotation is shown by the rhomboid structures, whose amplitudes are not uniform across all wavelengths. The vertical black dotted lines indicate the size of the features used to compute the summed fluxes shown in the four bottom panels (errors have the size of the symbols). Wavelengths are in in vacuum and in the laboratory reference frame.
}
\label{fig:absorption}
\end{figure}

\begin{figure}[h!]
\centering
\includegraphics[width=0.99\hsize]{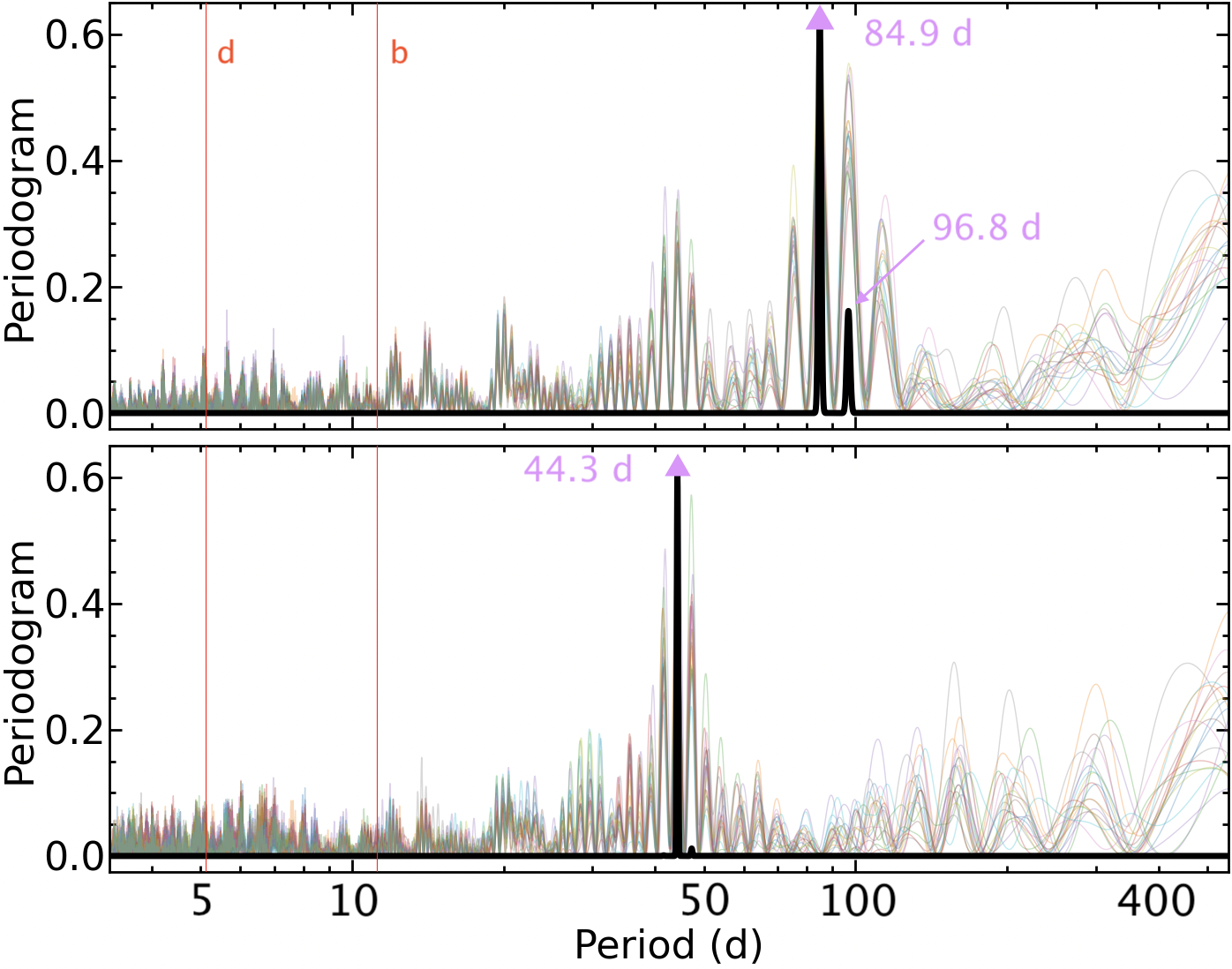}
\caption{Top panel shows the joint periodogram of photospheric features ("rhomboids") in thick black, with individual GLS periodograms of 20 rhomboid light curves overplotted in color. A significant peak appears at 84.9 $\pm$ 0.6 d (magenta triangle), with a second, weaker peak at 96.8 $\pm$ 0.7 d (magenta arrow). Bottom panel shows the joint periodogram of the residuals after removing the 84.9-d signal, with a peak at 44.3 $\pm$ 0.2 d (magenta triangle) and its weaker 2-yr alias. False alarm probabilities (0.1\,\%, 1\,\%, and 10\,\%) of the joint periodograms lie below 0.05 and are not visible in the diagrams. The vertical red lines denote the orbital periods of Proxima b and d.}
\label{fig:absorption_periodogram}
\end{figure}

\section{Stellar rotation period \label{sec:absorption}}

Plotting all 117 flattened differential spectra together reveals enhanced variability at specific wavelengths. The first and second panels of Figure~\ref{fig:absorption} show a narrow wavelength range of the spectra dominated by photospheric TiO transitions. In the second panel, corresponding to the differential spectra, rhomboid structures appear throughout. These features have typical widths of 0.07--0.10 \AA~and amplitudes of 13--100 ppt (approximately 2--10 $\sigma$ detections at red wavelengths). They are more easily detected at longer wavelengths, as the dispersion of the residual spectra increases toward the blue. In Figure~\ref{fig:absorption}, we selected four of the rhomboids and computed the sum of the differential fluxes in the narrow wavelength intervals marked by the vertical dotted lines. The resulting light curves are illustrated as a function of observing time in the bottom four panels of the figure. These light curves appear to be specular images of each other, depending on whether the feature traces the pseudo-continuum or the absorption, and all resemble the variability pattern of the full-width at half-maximum (FWHM) of the cross-correlation functions (CCFs)   discussed in \citet{mascareno20, mascareno25}.

The variability of most rhomboid structures reflects the photospheric rotation of Proxima Centauri. We extracted light curves randomly for a few tens of rhomboids at wavelengths longer than 6000 \AA~and computed their periodograms, which were analyzed jointly. The amplitudes were scaled to a reference feature, and individual Generalized Lomb–Scargle (GLS; \citealt{zechmeister09}) periodograms were computed over periods ranging from 3 to 550 d. The lower bound corresponds to twice the median ESPRESSO cadence, while the upper bound equals half the total temporal baseline of the ESPRESSO observations and is approximately five times the longest reported stellar rotation period ($\sim$112 d; \citealt{mascareno15}). We combined all features by multiplying the periodogram power at each frequency, enhancing coherent signals while suppressing noise. This approach, inspired by the stacked Bayesian GLS method of \citet{mortier17} (see also \citealt{stevenson23}), generates the joint periodogram shown in the top panel of Figure~\ref{fig:absorption_periodogram}. The False Alarm Probabilities (FAPs) at 0.1\.\%, 1\,\%, and 10\,\%~levels for the joint periodograms were computed using a bootstrap method, by generating 5,000 synthetic datasets for each index by randomly permuting the relative fluxes while preserving the observation times and the noise distribution.

In the top panel of Figure~\ref{fig:absorption_periodogram}, a significant peak is detected at 84.9\,$\pm$\,0.6\,d, where the uncertainty corresponds to the half width at half maximum of the peak. A second, weaker peak appears at 96.8\,$\pm$\,0.7 d and may have multiple interpretations. This signal could be related to the primary peak by a 2-yr alias, or it may represent an independent rotational modulation possibly caused by differential rotation. Assuming the two values correspond to the stellar equatorial and polar rotation periods, we estimated a rotation shear of $\Delta \Omega = 0.009 \pm 0.001$ rad\,d$^{-1}$ and a dimensionless differential rotation coefficient of $\alpha = 0.14 \pm 0.02$. Direct comparison with stars of similar rotation rate is not possible, as all low-mass stars with measured rotational shear rotate faster than Proxima Centauri \citep[e.g.,][]{reinhold13, zaleski20, araujo23, valio24} We nevertheless favored the alias interpretation, as the 96.7-d signal vanishes from the data simultaneously with the strongest peak and 84.9 d is more consistent with previous stellar rotation determinations (see below). Furthermore, the spectral window of the ESPRESSO time series (Figure~\ref{fig:spectral_window}) indicates that power from intrinsic periodicities may be redistributed to periods with aliases near 0.5 d, 1 d, and approximately 2 yr. 

The primary signal at 84.9\,$\pm$\,0.6 d is fully compatible with the independent value of 84.35\,$\pm$\,0.90 d reported by \citet{mascareno25} from Near Infra Red Planet Search (NIRPS) data using a Bayesian Gaussian process analysis. It also agrees with earlier photometric measurements in the interval 82.5--86.4 d based on {\sl Hubble} Fine Guidance Sensor (FGS) data \citep{benedict98}, all-sky ASAS photometry \citep{kiraga07, mascareno16, wargelin17, damasso20a}, and Skynet Robotic Telescope Network photometry \citep{engle23}. Using MEarth Survey photometry and spectropolarimetric data, \citet{newton18} and \citet{klein21} measured rotation periods of 89 d and 89.8 d, respectively, for Proxima Centauri, slightly longer than the values reported above. \citet{newton18} cautioned that their measurements may be affected by seasonal monsoon patterns. The observations of \citet{klein21} spanned only a single rotation cycle, which is insufficient to determine the stellar rotation period with high statistical significance. All of these studies focused on the photospheric variability of Proxima Centauri.

Motivated by the results of \citet{benedict98}, who identified two distinct behavior modes in their FGS data (higher-amplitude, longer-period and lower-amplitude, shorter-period), we removed the 84.9-d signal from each rhomboid light curve by fitting the following expression:
\begin{equation}
f = a + b\, x + c\, {\rm sin}\, (2\, \pi\, x + \phi)
\end{equation}
where $x$ is the phase (the fractional part of $t/P_{\rm rot}$, with $t$ denoting time) and the rotation period, $P_{\rm rot}$, is fixed at 84.9 d. The parameters $a$ (zero), $b$ (linear slope), $c$ (amplitude of the sinusoidal variation), and $\phi$ (phase correction) are free for each light curve. We then applied the same procedure as before to compute the joint periodogram of the residuals, shown in the bottom panel of Figure~\ref{fig:absorption_periodogram}. A significant peak is detected at 44.3\,$\pm$\,0.2 d. We note that the $\approx$47-d signal reported by \citet{mascareno20} from a Gaussian process analysis of the ESPRESSO CCF FWHM data corresponds to the 2-yr alias of our 44.3-d period. As also found by \citet{benedict98}, the 84.9-d signal generally exhibits larger amplitude variations than the 44.3-d signal. Although the latter is close to half the stellar rotation period, it is sufficiently offset to produce a distinct signature, consistent with the presence of active regions on the stellar surface located near 180$^\circ$ apart in longitude.

After removing the 44.3-d signal, a long period signal ($\sim$510 d) emerged near the upper limit of the joint periodogram, potentially linked to the 442-d activity cycle reported by \citet{cincunegui07}. More recent long-term studies pointed to a longer multi-year magnetic cycle for Proxima Centauri (6.5--8 yr; \citealt{mascareno16, wargelin17, damasso20a, azizi24, wargelin24}, and $\sim$18 yr, \citealt{mascareno25}), which is not fully covered by the ESPRESSO observations. We emphasize that after three iterations, no photospheric variability is detected at the orbital periods of Proxima b and d.

\begin{figure*}[h!]
\sidecaption
\includegraphics[width=12.9cm]{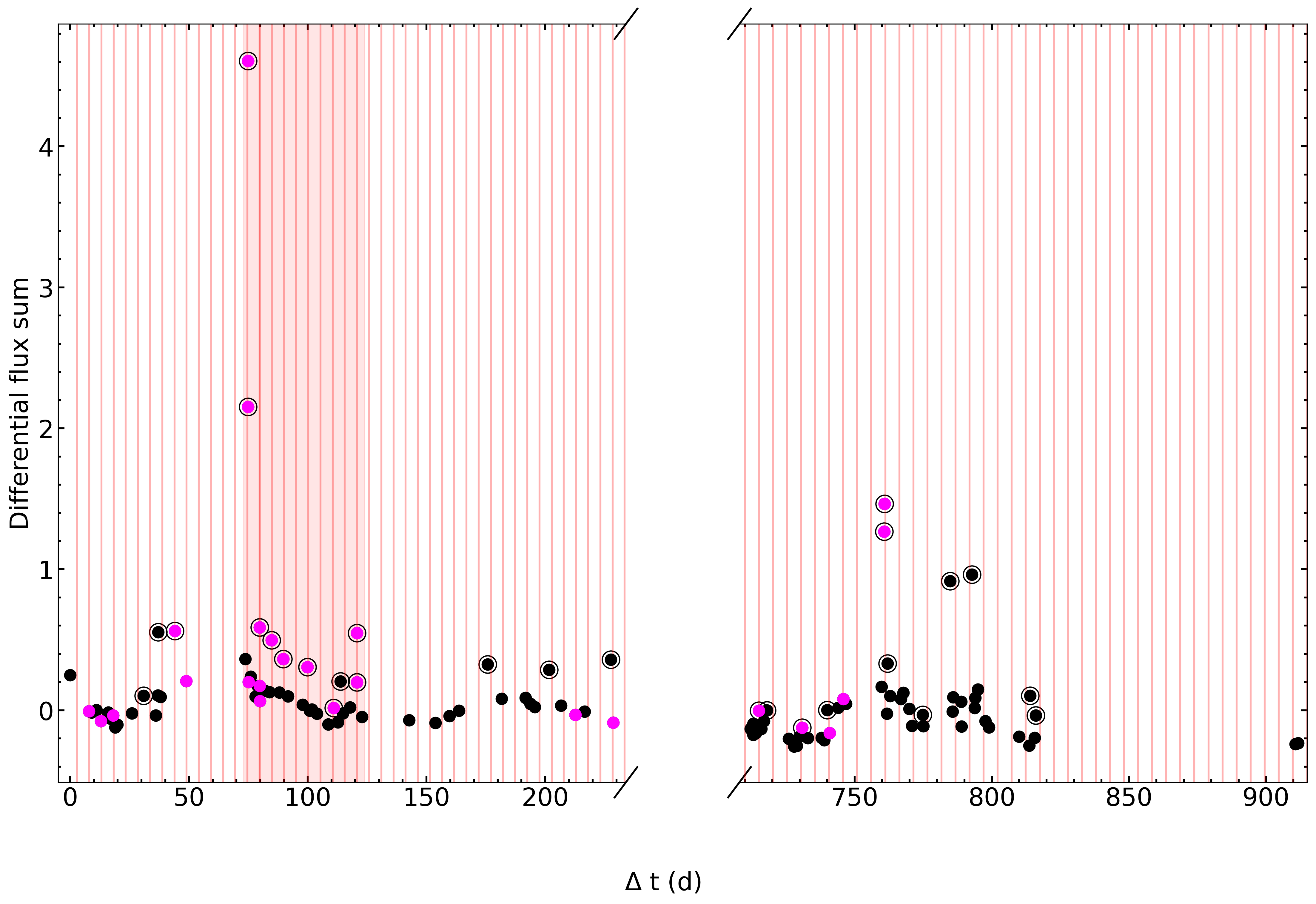}
\caption{The light curve of the Fe\,{\sc i} emission line at 5457.125~\AA\ is shown as circles. The vertical red lines (predicted $t_o$) are spaced at the orbital period of Proxima~d ($P_{\rm orb}=5.12338$ d), and the reddish band highlights the highest cadence interval (2019 April 26 -- June 13). Encircled symbols indicate epochs flagged with enhanced iron emission, while magenta symbols denote epochs within $\pm$0.45 d of the nearest red lines (see text). The most recent epoch is omitted for clarity.}
\label{fig:d}
\end{figure*}

\section{Likely star--planet interaction \label{sec:spi}}

\citet{cuntz00} presented a study on the interactions between giant exoplanets and their host stars, suggesting that such interactions can enhance stellar winds as well as coronal and chromospheric activity. They proposed two mechanisms, tidal interactions and direct magnetic coupling, scaling with the star--planet distance as $a^{-3}$ and $a^{-2}$, respectively. Magnetic activity should coincide with the planetary orbital period $P_{\rm orb}$, whereas tidally induced activity is expected to vary with a period of $P_{\rm orb}/2$ as the tidal potential exerted by the planet on the star has two maxima per orbital cycle. Because Proxima b and d orbit very close to their host star, they are likely subject to strong tidal forces that can deform their interiors and generate internal heating, although significant tidal evolution of Proxima b’s orbit appears unlikely \citep[e.g.,][]{ribas16}. In addition, the two planets are separated by just 0.02 au, inducing strong gravitational interactions between them \citep{livesey24}. Proxima b and d are probably rocky and may possess iron-rich, electrically conducting cores capable of sustaining convection-driven dynamos and magnetic fields \citep[e.g.,][]{stevenson03}. Dynamo models indicate that surface magnetic field strengths vary only modestly for Earth-mass planets and increase slightly for moderately more massive super-Earths \citep{driscoll11}. For rapidly rotating planets, field strength scales with the convective energy flux \citep{christensen09, christensen10}, while core-mantle boundary heat-flow constraints suggest that more massive super-Earths may host shorter-lived dynamos \citep{tachinami11}. In the Proxima Centauri system, Proxima b is roughly four times more massive than Proxima d but orbits farther from the star and receives about three times less stellar irradiation. Proxima d’s closer orbit therefore makes it a particularly promising candidate for detectable star--planet interactions.

\subsection{Proxima d \label{sec:d}}

Given the frequent flaring of Proxima Centauri, we conducted a statistical analysis to assess the significance of potential planet-induced emission in the ESPRESSO data. Part of our ESPRESSO observations overlap with {\sl TESS} Sectors 11, 12, and 38. Except for 2019 April 26, none of the 29 ESPRESSO epochs contemporaneous with {\sl TESS} coincide with the peaks or fall within prominent photometric flares. The strongest ESPRESSO event on 2019 April 26 appears rather faint in the {\sl TESS} light curve (Figure~\ref{fig:tess}), with an amplitude comparable to the weakest flares reported by \citet{vida19}. We therefore defined an empirical criterion to identify  "flares'' (or enhanced emission) in the ESPRESSO Fe\,{\sc i} $\lambda$5457.125~\AA\ data: an epoch is flagged if its index exceeds the local mean by more than 3\,$\sigma$, or if it follows an observation already classified as enhanced emission and is much brighter than the local mean. The first and last three epochs were excluded due to the lack of adjacent measurements or large temporal gaps. We adopted the Fe\,{\sc i} lines because they exhibit lower stochastic variability than other chromospheric lines, enabling a more robust estimation of the local mean. Using this definition, we identified 28 enhanced emission epochs, which implies that $24.8 \pm 4.7$\,\%~of the observing time was classified as "flaring". This fraction is substantially higher than the 7.2\,\%~reported by \citet{vida19} from {\sl TESS} observations. Using other Fe\,{\sc i} lines (Figure~\ref{fig:iron}) or adopting a different 5\,$\sigma$ threshold for the enhanced-emission flag does not alter the conclusions of this work. The higher threshold filters out an equal number of Proxima d aligned and non-aligned epochs from the enhanced-emission designation.

Figure~\ref{fig:d} presents the light curve of Fe\,{\sc i} $\lambda$5457.125~\AA\ line\footnote{All wavelengths mentioned in the text and shown in the figures are in vacuum and laboratory rest frame.}, highlighting the epochs of  enhanced and non-enhanced iron emission. The amplitude of the neutral iron enhancements varied on timescales comparable to the stellar rotation, with many Fe\,{\sc i} events reaching flux levels $\ge$25\,\%~above the local mean.  Figure~\ref{fig:d} also shows vertical markers spaced by the orbital period of Proxima d ($P_{\rm orb}=5.12338 \pm 0.00035$ d; \citealt{mascareno25}) and referenced to the epoch $t_o = 2458604.61011$ BJD ($\Delta t \approx 80$ d in the figure). Epochs within $\pm$0.45 d ($\pm$0.088 in Proxima d's phase) of the predicted $t_o$ across the full ESPRESSO dataset are clearly identified in the Figure and form the basis of the following statistical analysis. 

We found a possible correlation between enhanced neutral iron emission and Proxima d. During the interval from 2019 April 26 to June 13 (highlighted in Figure~\ref{fig:d}), which spans 49.17 d, slightly more than half a stellar rotation cycle and corresponding to the highest ESPRESSO cadence, Proxima d completed 9.6 orbital cycles. On seven of these occasions, there are 12 ESPRESSO observations obtained within $\pm$0.45 d of the predicted $t_o$, nine of which coincide with enhanced neutral iron emission from Proxima Centauri, including the strong flare on 2019 April 26. This yields a 75\,\%~temporal coincidence between enhanced Fe\,{\sc i} emission events and the orbital period of Proxima d, which is very suggestive. Outside this interval, the coincidence rate drops to 38\,\%. This pattern was consistently observed in all seven Fe\,{\sc i} lines between 5329--5458 \AA\ (Figure~\ref{fig:iron}) and in all other chromospheric species (e.g., Figures~\ref{fig:caii} and~\ref{fig:nai}), though it is less clearly visible in the latter due to larger flare amplitudes, possible flare phase shifts (delays) between species, and inadequate event sampling.

In Figure~\ref{fig:d} there are $k = 11$ independent flares phased with the planet (intra-night, consecutive flare-flagged epochs are counted as a single event) out of a total of $n = 25$ single flares. The ratio $p = 0.176$ between the adopted phase tolerance (2$\times$0.45 d) and the  planet’s orbital period defines the expected flare frequency under the null hypothesis of random distribution of flares with planetary phase. The binomial probability mass function:
\begin{equation}
f = \binom{n}{k} ~p^{k} ~(1 - p)^{n - k}
\end{equation}
yields $f = 0.14$\,\%, suggesting that the observed clustering of flares in phase space is unlikely to arise from random timing, and supporting a non-random, potentially physical association with the planetary orbit (flare--planet relation). Similarly, to quantify the probability of observing Proxima Centauri in enhanced iron emission at the same orbital phase, we used the enhanced-emission frequency $p = 0.248$, $k = 14$ enhanced-emission epochs within  $\pm$0.45 d of the planetary phase out of $n = 25$ epochs with similar orbital configuration, we obtained  $f = 0.064$\,\%. This result disfavors a purely random origin and indicates a possible physical link between enhanced iron emission and the planet’s orbital configuration (planet--flare relation). Overall, the data points to a bijective flare–planet and planet–flare association with a probability of 99.9\,\%~in the ESPRESSO observations.

\begin{figure*}[h!]
\centering
\includegraphics[width=0.33\hsize]{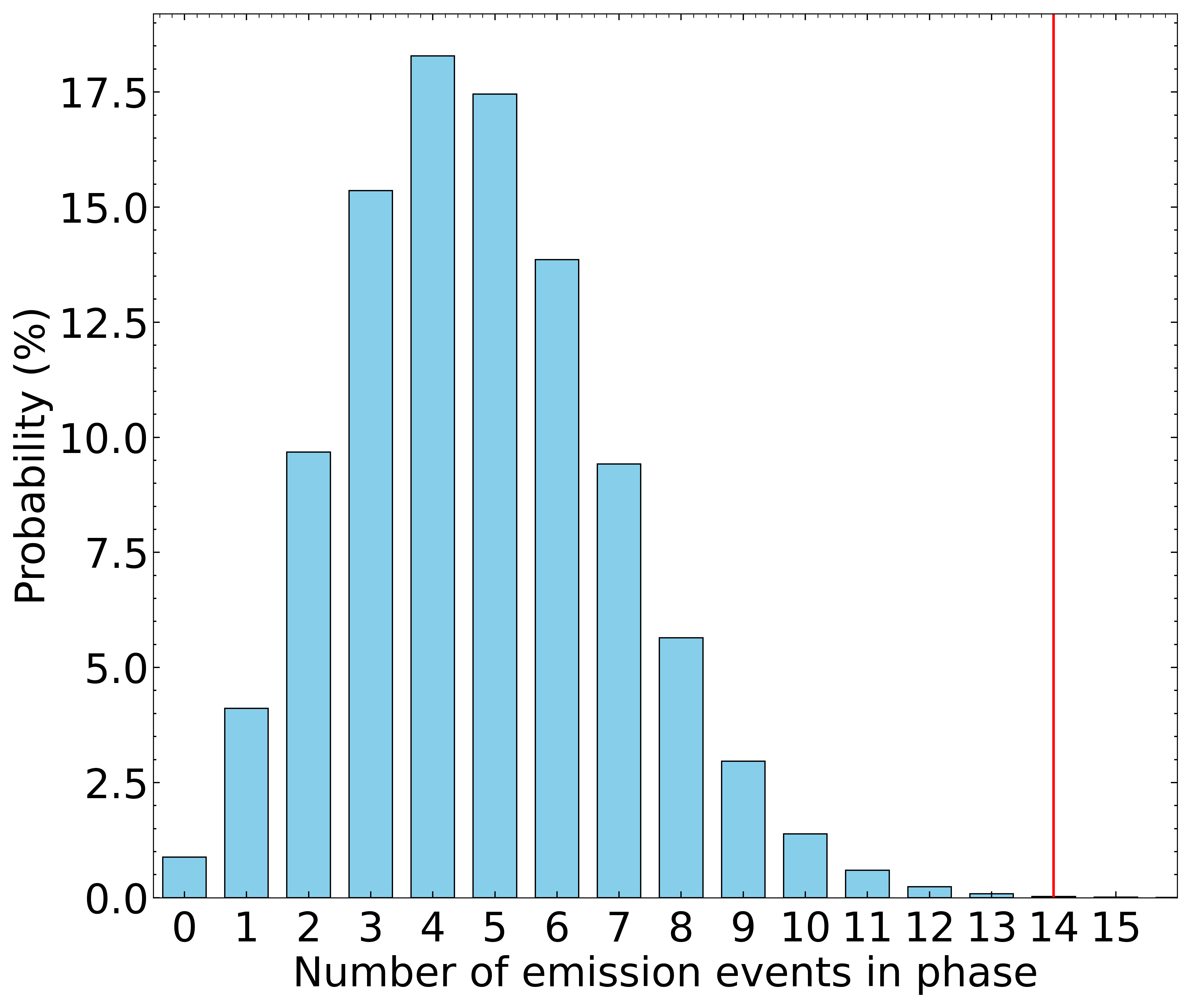}
\includegraphics[width=0.33\hsize]{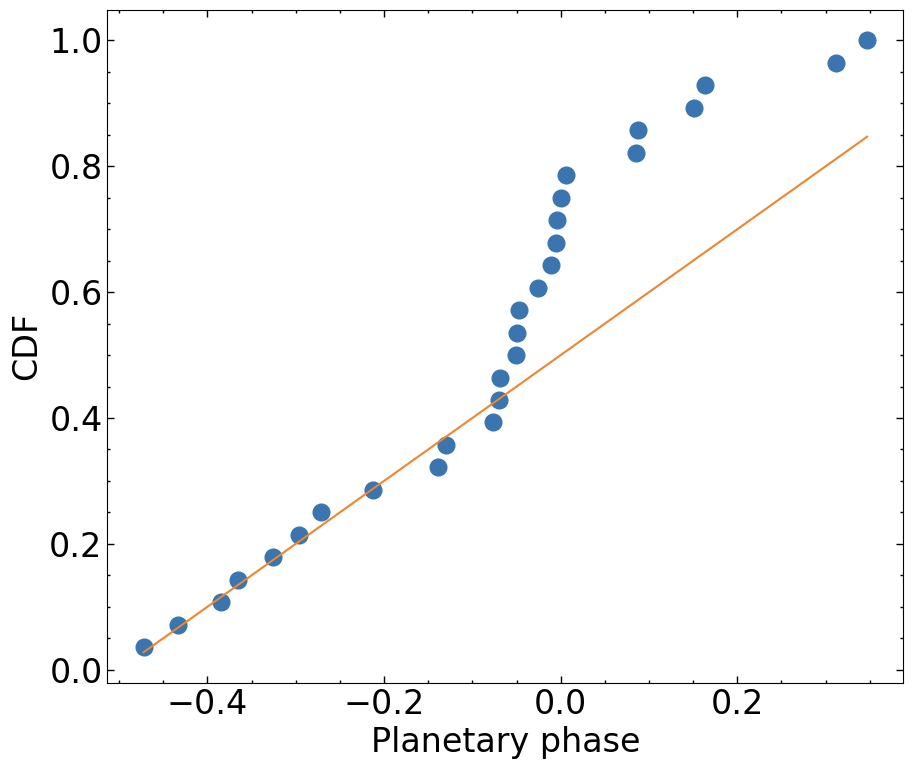}
\includegraphics[width=0.33\hsize]{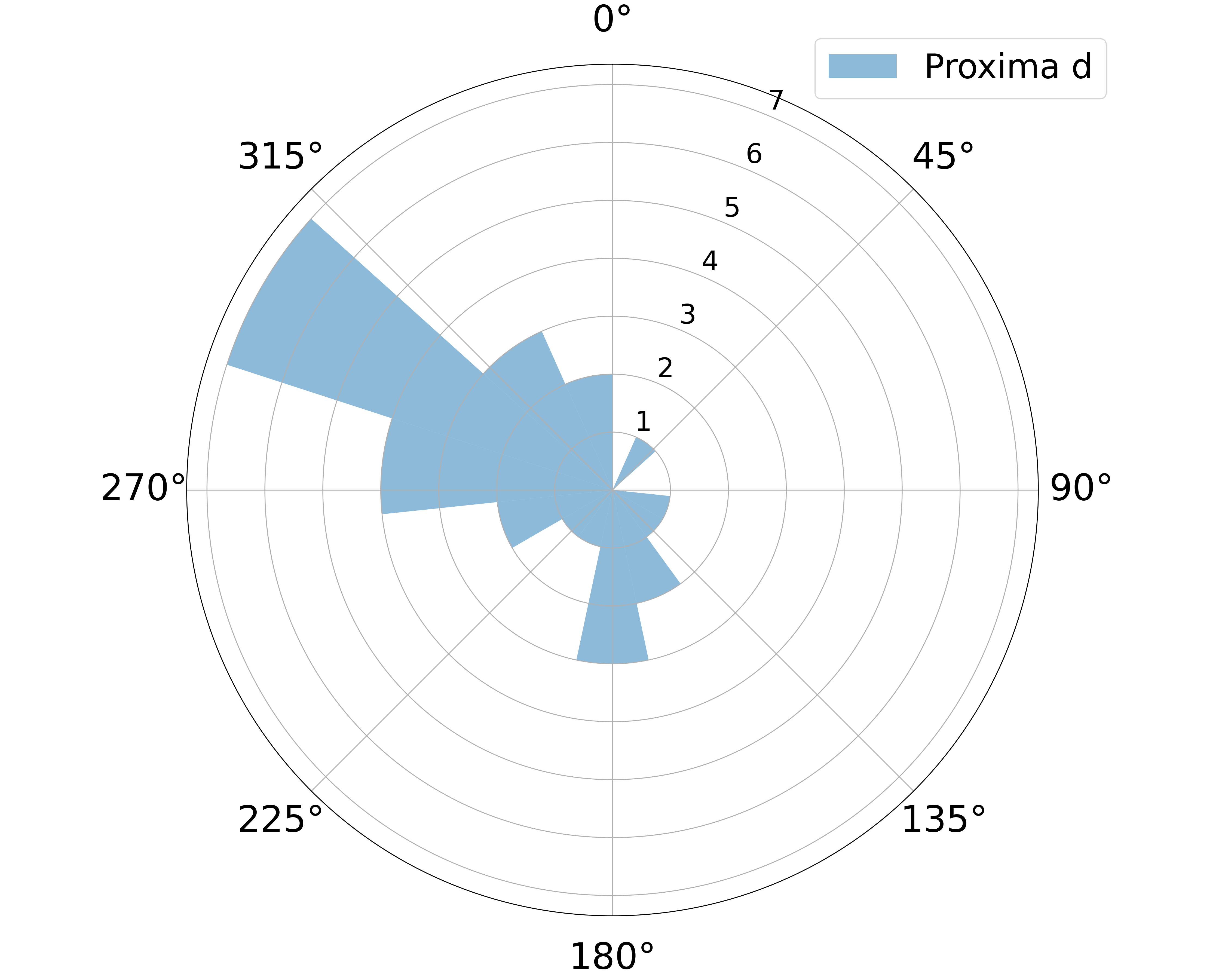}
\caption{Statistical tests for Proxima d. {\sl (Left)} Probability distribution, from Monte Carlo simulations, of the number of enhanced Fe\,{\sc i} $\lambda$5457.125 \AA~emission events in the ESPRESSO data that are phased with the orbital period of Proxima d. The red vertical line marks the actual number of events in the ESPRESSO data that coincide with the orbital phase of Proxima d.
{\sl (Middle)} Cumulative distribution function of Fe\,{\sc i} $\lambda$5457.125~\AA\ enhanced-emission epochs as a function of Proxima d phase (circles, zero phase = 2458604.61011). The straight line represents the expected cumulative distribution of flares if they were uniformly distributed across all phases.
{\sl (Right)} Histogram (polar plot) of the 28 enhanced iron emission epochs, folded in phase using Proxima d’s orbital period. Bin size is 22\degr5. The observer’s direction is at 0\degr, and orbital motion is clockwise. Note that for this plot we used the zero phase published by \citet{mascareno25}, which has an uncertainty of 11\degr~(non-transiting planet).
}
\label{fig:all_tests_d}
\end{figure*}

To further assess this result, we performed Monte Carlo simulations with 10$^6$ trials to estimate the probability of obtaining $m = 0, 1, 2, \ldots$ enhanced-emission epochs within $\pm$0.45 d of the predicted planetary phase, considering the ESPRESSO effective observing window (324.84 d) and the total number of epochs (113) in this window. To increase the realism of the simulations, both the flare frequency $p = 0.248 \pm 0.047$ and the orbital period of Proxima~d were drawn from normal distributions centered on their adopted values, with standard deviations corresponding to the 1$\sigma$ uncertainties. The minimum separation between simulated observing epochs was fixed at 0.5 h, accounting for the ESPRESSO exposure time and overheads. Left panel of Figure~\ref{fig:all_tests_d} shows the probability distribution from the Monte Carlo simulations. The probability of observing 14 enhanced-emission events in phase with Proxima~d in the ESPRESSO dataset is 0.03\,\%, indicating that such a clustering by chance is quite unlikely. To account for the exact arrangement of ESPRESSO observing times, including the high-cadence interval and the large time gaps between observing seasons, we performed 10$^6$ Monte Carlo permutation simulations using the 113 ESPRESSO epochs. In each simulation, the enhanced-emission flag was randomly reassigned to 28 epochs, preserving the total number of flagged emission events. The simulations yielded a 0.01\,\%~probability of observing 14 enhanced-emission events in phase with Proxima d by chance (see Figure~\ref{fig:mc_permutation}), consistent with previous results.

Changing the phase tolerance does not affect the main conclusions of this work. On one hand, reducing the phase tolerance decreases the number of events aligned with the planet, yet all statistical tests indicated that the number of events phased with the orbital period remains too high to be explained by randomness, with probabilities fluctuating between 0.01\,\%~and 3\,\%, likely due to the small sample size. On the other hand, increasing the phase tolerance slightly increases the number of aligned events, and the corresponding Monte Carlo probabilities also rise due to the high intrinsic flare frequency. A tolerance of $\pm$0.45 d was adopted based on the width of the enhanced-emission phase distribution (see below) and represents a compromise between ensuring that the number of events likely aligned with the planet is statistically significant ($>$10) while remaining small enough to avoid including many random events. A phase tolerance is probably needed to account for potential offsets of the subplanetary point \citep{shkolnik03}, uncertainties in the orbital ephemeris, stellar differential rotation, changes in the geometry and size of stellar active regions, and/or temporal variations in the stellar magnetic field configuration.

Following the approach of \citet{ilin22} and \citet{ilin24}, we also applied the Anderson--Darling test to the ESPRESSO enhanced iron emission data to assess the significance of the star--planet interaction signal \citep{anderson52}. Specifically, we tested whether the cumulative distribution function (CDF) of the phase-folded emission data is consistent with a uniform distribution across orbital phases (middle panel of Figure~\ref{fig:all_tests_d}). The observed CDF shows a sharp rise near phase zero (the test is independent of the exact phase at which the rise occurs). The test evaluated whether the residuals (observations minus the expected uniform distribution) are consistent with a normal distribution. We obtained a $p$-value of 0.00192, indicating that the hypothesis of flare-distribution uniformity can be rejected at 99.8\,\%~confidence. Right panel of Figure~\ref{fig:all_tests_d} shows a polar plot of all 28 enhanced emission epochs folded in phase with Proxima d's orbital period, providing another illustration of the accumulation of stellar flares at a similar orbital phase. Fitting a Gaussian function to the phase distribution yielded a phase width of $0.082 \pm 0.018$ ($29.5 \pm 6.5$ deg), corresponding to $0.42 \pm 0.09$ d. For this figure, we adopted $t_o = 2460557.55 \pm 0.16$ as the zero phase, corresponding to the inferior conjunction from the radial velocity solution of \citet{mascareno25}. The large uncertainty arises from the planet’s non-transiting nature. Our adopted zero phase differs from their value by 0.93 d. We note that our zero phase is anchored to one of the enhanced emission epochs and therefore does not necessarily coincide exactly with the time of inferior conjunction; however, both epochs must be close, as the flares must occur on the visible hemisphere of the host star to be observable.

\subsection{Proxima d magnetic field strength}

All statistical evidence points to a likely connection between Proxima d and enhanced emission correlated with the planetary orbital period. However, not all observed events can be attributed to the planet (see Figure~\ref{fig:d}), and the planetary passage through the visible stellar hemisphere does not always produce such enhanced emission. This periodicity favors a magnetic star--planet interaction rather than a tidal origin. Additionally, our results, together with the predominantly poloidal and moderately axisymmetric magnetic field of Proxima Centauri \citep{klein21}, are consistent with the scenario proposed by \cite{lanza12}, in which such fields can induce flaring activity phase-locked to the planet.

Proxima d orbits at $\approx$44 times the star's radius (R$_*$) in a nearly circular orbit. The Alfv\'en radius of Proxima Centauri is estimated to be $\sim$25 R$_*$ (\citealt{klein21}; see also \citealt{kavanagh21}), although it may extend up to 46.5 R$_*$ during periods of maximum stellar activity \citep{alvarado20}. Magnetohydrodynamic simulations further predict that Proxima b and d are subject to strong and variable stellar wind pressures, with transitions between sub- and super-Alfv\'enic regimes that may occurr on timescales as short as one day \citep{garraffo22}. Proxima d may therefore reside in the sub-Alfv\'enic regime intermittently, over both short and extended intervals. Our ESPRESSO observations show small and modest phase drifts in the enhanced emission events correlated with the planet, together with episodic stronger clustering (e.g., between 2019 April 26 and June 13), which may be more consistent with the super-Alfv\'enic stretch-and-break mechanism (helicity-driven reconnection) proposed by \citet{lanza12}. Nevertheless, alternative scenarios cannot be excluded specially because Proxima Centauri was transitioning out of its activity maximum during the ESPRESSO observations \citep{savanov12, mascareno25}. Proxima d may then hold Alfv\'en waves that propagate along magnetic field lines and deposit energy at stellar magnetic foot points, producing chromospheric emission that co-rotates with the planet \citep[e.g.,][]{ip04,strugarek19}. Additionally, \citet{fischer19} suggested that, in the presence of a stellar magnetic anomaly within the passageway of the magnetic foot points, an Alfv\'en wing could trigger multiple nanoflares along its trajectory, leading to emission pulsed with the planetary orbital period under quasi-continuous flare triggering.

Using Equation~1 of \citet{vida19}, we found that the relative energies of the enhanced neutral iron emission events in the ESPRESSO data span a factor of $\approx$60. In the {\sl TESS} band (0.6--1.0 $\mu$m), the most energetic ESPRESSO event corresponds to a total radiated energy of $\approx 10^{30}$ erg \citep{vida19} and a duration of 2500 s. This implies flare luminosities of $6.5\times10^{24}$ – $4\times10^{26}$ erg\,s$^{-1}$ for events of comparable duration in the ESPRESSO data. We used these luminosities to estimate the strength of the planetary magnetic field using different analytical predictions for the power released in magnetic star--planet interaction. The large-scale magnetic field strength of Proxima Centauri has recently been measured to be about 200 G \citep{klein21}, whereas earlier studies reported stronger values in the range 450--2000 G \citep{guedel04, reiners08}. We estimated the expected star--planet interaction luminosity from Equations 2 and 6--9 from \citet{ilin24}, which build on the formulations and approximations of \citet{lanza12}, \citet{saur13}, and \citet{kavanagh22}. We adopted an average polar magnetic field of 600 G for Proxima Centauri, a Mars-sized Proxima d, an average coronal density of $4 \times 10^{11}$ cm$^{-3}$ \citep{guedel04}, which is 40 times higher than the solar value ($2 \times 10^{10}$ cm$^{-3}$; \citealt{young09}), and the remaining parameters listed in Table~\ref{tab:parameters}. For an unmagnetized planet, the resulting star--planet interaction luminosity is $10^{24}$ erg\,s$^{-1}$ in the stretch-and-brake scenario and $\leq 10^{21}$ erg\,s$^{-1}$ in the Alfv\'en wing scenario. Both predictions underestimate the observed luminosity, although the stretch-and-brake case yields a closer value. This may indicate the presence of an intrinsic magnetic field in Proxima d. Assuming a magnetized planet, we infer a polar field strength of the order of $\sim10^{4}$ G in both interaction regimes. Adopting instead a stellar magnetic field strength of 2000 G \citep{guedel04} reduces the required planetary field to $\sim10^{3}$ G. In any case, these values exceed the stellar surface field and the estimates available for giant exoplanets \citep{cauley19}, therefore appearing physically implausible

To account for higher luminosities, other formalisms must be invoked. \cite{lanza13} further explored the stretch-and-brake scenario by evaluating the Poynting flux through the base of a magnetic flux tube connecting the planetary and stellar surfaces, obtaining the expression:
\begin{equation}
P_{\rm SPI}  \approx \frac{2\,\pi}{\mu_0} \, f_{\rm AP} \, R_{\rm p}^2 \, B_{\rm p}^2 \, v_{\rm rel}
\end{equation}
where $P_{\rm SPI}$ is the emitted power (luminosity), $\mu_0$ the vacuum magnetic permeability, $R_{\rm p}$ the planetary radius, $v_{\rm rel}$ the relative velocity between the planet and stellar magnetic field lines at the orbital distance, $B_{\rm p}$ the polar planetary magnetic field strength, and $f_{\rm AP}$ the fraction of the planetary surface magnetically connected to the stellar field. The latter depends on the ratio of the stellar magnetic field at the planetary orbit to the planetary field strength as follows \citep{adams11}:
\begin{equation}
f_{\rm AP} = 1 - \left( 1 - \frac{3 \, \beta^{1/3}}{2 + \beta} \right)^{1/2}
\end{equation}
where $\beta = B_*(a) / B_{\rm p}$. The stellar magnetic field at the planetary orbit $a$ is $B_*(a) = B_*^o (R_*/a)^s$, with $B_*^o$ the polar stellar field strength and $2 \le s \le 3$ depending on the magnetic field geometry \citep{lanza13}: $s=2$ for a radial field and $s=3$ for a dipolar configuration. We solved these equations numerically and derived a magnetic field strength for Proxima d in the range 3--280 G. This broad interval accounts for both stellar field geometries, the plausible 200--2000 G range for Proxima Centauri, the observed flare luminosities, and two possible planetary radii for Proxima d (Earth and Mars sizes). Table~\ref{tab:magfield} provides estimates for individual cases, systematically exploring the parameter space. Among the physical parameters, flare luminosity has the largest impact, followed by stellar field geometry and planetary radius. For an intermediate geometry ($s=2.5$), a stellar field of 600 G, mean flare luminosity of $10^{25}$ erg\,s$^{-1}$, and Mars radius, the inferred median polar field strength is 16.4 G and about half as large if Proxima d has an Earth-size radius. This value implies that approximately $f_{\rm AP} = $ 11\,\%~of the planetary surface is threaded by stellar magnetic field lines associated with the interconnecting loop. For comparison, Earth’s polar field is 0.6 G \citep{alken21}, while Jupiter’s reaches $\sim$50 G \citep{connerney18}. A polar field of 16 G is high, yet theoretically plausible for a hot terrestrial planet \citep{zuluaga12}. Such a value would also imply strong core heat flux \citep{christensen09}. However, these estimates depend on poorly constrained parameters, including the planetary radius and the stellar magnetic field, rendering the planet's magnetic field strength determination reliable only at the order-of-magnitude level.

\citet{ilin22} searched for flaring star--planet interaction in the M-dwarf AU Mic using {\sl TESS} optical light curves of its 11.7‑M$_\oplus$ planet, AU Mic b, but found no evidence of planet-induced flares. On the contrary, using high-resolution spectroscopy, \citet{klein22} reported a 3 $\sigma$ modulation of the He\,{\sc i} D3 emission flux at the orbital period of AU Mic b. This and our results suggest that flares of M-type stars triggered by the innermost planets may be low in energy and too weak to detect at red wavelengths. The counterexample is HIP\,67522, a very young star for which \citet{ilin25} found that 15 stellar flares observed with {\sl TESS} and {\sl CHEOPS} over a period of 5 yr cluster near the transit phase of the innermost planet ($\sim$14 M$_\oplus$, \citealt{thao24}). Note that HIP\,67522 is a 17 Myr old, G-type star (1.22 M$_\odot$; \citealt{rizzuto20}), while Proxima Centauri is about 300--400 times older and 10 times less massive. 
 \citet{pineda23} detected coherent radio bursts from the slowly rotating M dwarf YZ Ceti at orbital phases corresponding to YZ Ceti b ($M_{\rm p}$\,sin\,$i = 0.70$ M$_\oplus$). If flare-triggering is confirmed to be caused by a magnetized planet, Proxima d ($M_{\rm p}$\,sin\,$i = 0.26$ M$_\oplus$) would become the smallest exoplanet known to host an intrinsic magnetic field capable of magnetospheric interaction with its host star, potentially contributing to enhanced stellar magnetic activity. In the solar system, Mercury ($\sim$0.38 M$_\oplus$) and Ganymede ($\sim$0.025 M$_\oplus$) are the smallest bodies known to host self-sustained magnetic fields \citep{ness75, kivelson96, breuer10}. In addition, this would have important implications for atmospheric (and possibly habitability) studies of Proxima d, as the planet (although on a non-habitable orbit) may host a magnetopause large enough to shield its surface from the stellar wind \citep[e.g.,][]{pena24}.

\addtolength{\tabcolsep}{0em}
\begin{table}[h!]
  \caption{\label{tab:parameters}Parameters used in the star--planet interaction equations of \citet{lanza13} and \citet{ilin24}.}
  \small
\centering
\setlength{\tabcolsep}{1.5pt}
\begin{tabular}{lcccl}
  \hline\hline\noalign{\smallskip}
 Parameter   & Proxima d & Proxima b  &  Ref$.$ & Description \\
  \hline
  \noalign{\smallskip}

$a$ (au) & 0.02881 & 0.04848 & 1 & Semimajor axis \\
$P_{\rm orb}$ (d) & 5.12338 & 11.18465 & 1 & Orbital period \\
$R_{\rm p}$ (R$_\oplus$) & 0.53 & 1.1 & 2 & Planetary radius \\
$P_{\rm rot}$ (d) & \multicolumn{2}{c}{84.9} & 2 & Stellar rotation period \\
$R_*$ (R$_\odot$) & \multicolumn{2}{c}{0.141} & 1 & Stellar radius \\
$B_*ô$ (G) & \multicolumn{2}{c}{600} & 2 & Stellar magnetic field  \\
$n$ & \multicolumn{2}{c}{0.10} & 3 & Constant $0 < n < 1$ \\
$\lambda^2$ & \multicolumn{2}{c}{1.29} & 3 &  Constant, eigenvalue \\
$\mu_{\rm 0}$ (H\,m$^{-1}$) & \multicolumn{2}{c}{4\,$\pi \times 10^{-7}$} & 4 & Magnetic permeability \\
$\Theta$ (deg) &  \multicolumn{2}{c}{90} & 2 & Wind--planet velocity angle  \\
$P_{\rm SPI}$ (erg\,s$^{-1}$) & \multicolumn{2}{c}{$10^{25}$} & 2 & Star--planet interaction power \\
$\rho_*$ (cm$^{-3}$) & \multicolumn{2}{c}{$4 \times 11^{11}$} & 5 & Coronal density \\

\noalign{\smallskip}
  \hline
\end{tabular}
\tablefoot{ 1.- \citet{mascareno25}. 2.- This paper (either measured from ESPRESSO data or adopted). 3.- \citet{ilin24}. 4.- \citet{CODATA}. 5.- \citet{guedel04}.
}
\end{table}

\begin{figure}[h!]
\centering
\includegraphics[width=0.75\hsize]{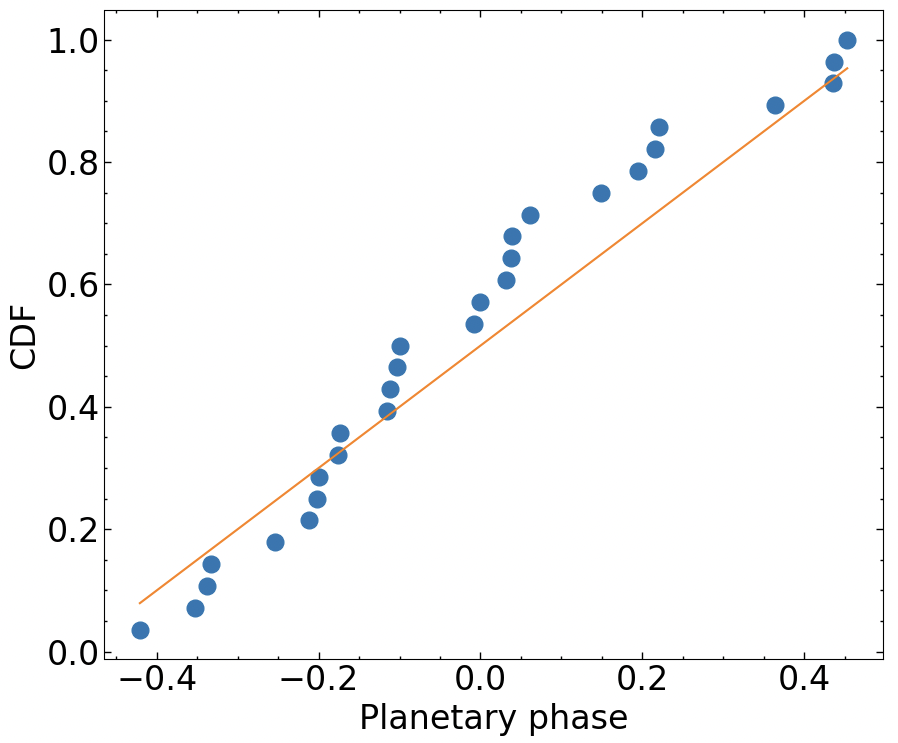}
\includegraphics[width=0.8\hsize]{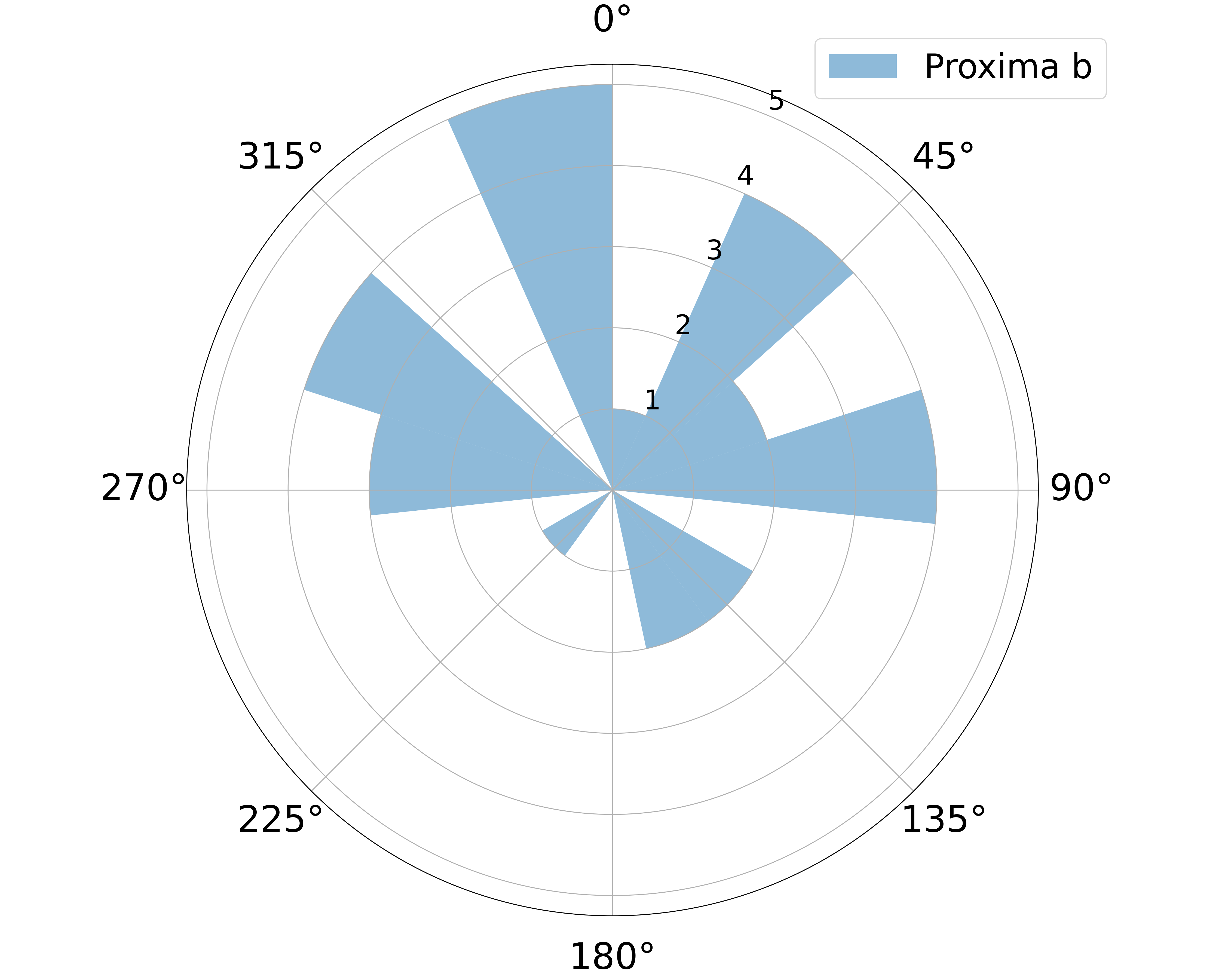}
\caption{Statistical tests for Proxima b. {\sl (Top}) Same as Figure~\ref{fig:all_tests_d} (middle panel) for Proxima b. {\sl (Bottom)} Same as Figure~\ref{fig:all_tests_d} (right panel) for Proxima b.
}
\label{fig:anderson-darling_b}
\end{figure}

\begin{figure}[h!]
\centering
\includegraphics[width=0.99\hsize]{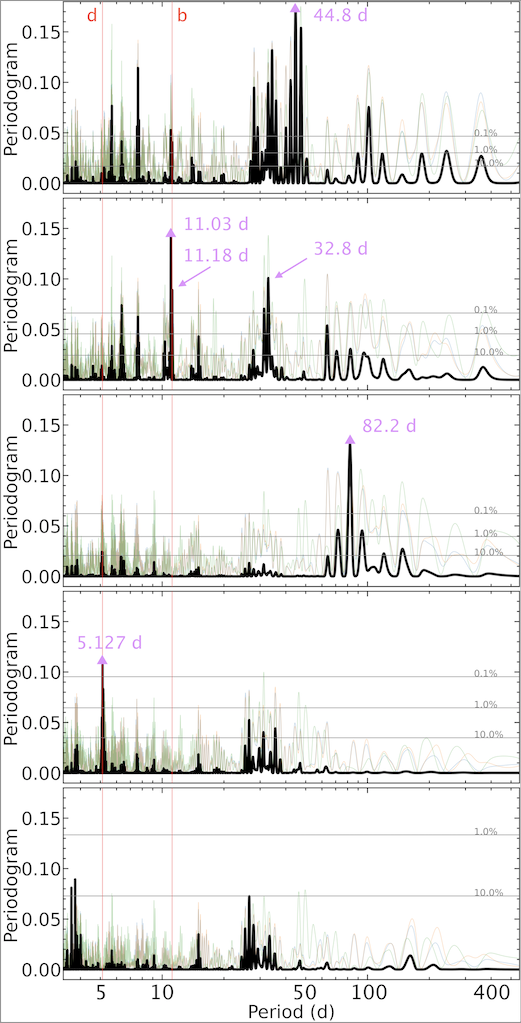}
\caption{Overplotted color periodograms of the Ca\,{\sc ii} H\&K and H$\alpha$ activity indices of Proxima Centauri, with the joint periodogram shown in black. The top panel displays the periodogram of the original differential flux sums; subsequent panels show periodograms of data residuals after sequential removal of the strongest signal marked by a magenta triangle. Periods mentioned in the text are  indicated.  FAP levels are shown by the horizontal gray lines. The vertical red lines denote the orbital periods of Proxima b and d.}
\label{fig:periodogram_b}
\end{figure}

\subsection{Proxima b}

We built the CDF of the ESPRESSO enhanced iron emission epochs phase folded on the orbital period of Proxima b ($P_{\rm orb}=11.18465 \pm 0.00053$ d; \citealt{mascareno25}), as shown in Figure~\ref{fig:anderson-darling_b} (top), and performed an Anderson--Darling test. The resulting $p$-value of 0.56 indicates no statistically significant correlation between the emission epochs and the orbital phase of Proxima b, suggesting that the planet is unlikely to trigger flares within the energy range probed by ESPRESSO. This is consistent with \citet{ilin24}, who reported similar results for the more energetic flares observed by {\sl TESS}. The polar distribution of the 28 enhanced emission epochs (Figure~\ref{fig:anderson-darling_b}, bottom) shows no preferred phase, consistent with the Anderson-Darling test, although most events occurred when the planet faced the visible stellar hemisphere. Figure~\ref{fig:cartoon} illustrates the orbital positions of Proxima b and d for each single enhanced emission event. We estimated the flare luminosity expected from star--planet interaction with an unmagnetized Proxima b (Table~\ref{tab:parameters}) following \citet{ilin24}, obtaining values $\le10^{24}$ erg s$^{-1}$, i.e., below our detection threshold. 

 \citet{shkolnik03} pioneered the search for planet-induced chromospheric variability/modulation using the Ca\,{\sc ii} H \& K lines, the strongest optical chromospheric diagnostics, together with H$\alpha$. According to these authors, magnetic heating of the star by the planet can be detected as stellar chromospheric activity modulated with the orbital period being more intense at phase 0 (subplanetary point). Figure~\ref{fig:caii} illustrates the variability of the Ca\,{\sc ii} H \& K and H$\epsilon$ lines Proxima Centauri. The three emission lines have very similar light curves, but they differ from the variability of the photospheric features (Figure~\ref{fig:absorption}).

Following the procedure described in Section~\ref{sec:absorption}, we computed the joint periodogram of all 117 Ca\,{\sc ii} H \& K (3969.591 \AA, 3934.777 \AA) and H$\alpha$ (6564.61 \AA) differential flux sums. No significant signal is detected at the stellar rotation period; instead, a forest of peaks appear at 44.8 $\pm$ 0.6 d (Figure~\ref{fig:periodogram_b}, top panel). This value is consistent, within 2\,$\sigma$ uncertainties, with half of the stellar rotation period derived in Section~\ref{sec:absorption}, indicating comparable rotational modulation in the photosphere and chromosphere. We excluded the strongest flare (2019 April 26) from the analysis to avoid any bias that the very high differential flux sums could introduce; however, we note that including it does not alter the results. The subsequent panels of Figure~\ref{fig:periodogram_b} display the individual and joint periodograms after sequential removal of the dominant signals at each iteration, following a prewhitening approach. After subtraction of the 44.8-d signal, the highest peak in the joint periodogram (second panel) occurs at 11.03 $\pm$ 0.03 d and exceeds the 0.1\,\%~FAP threshold. The third iteration (third panel) yields a signal at 82.2 $\pm$ 1.6 d. In the fourth iteration (fourth panel), the only peak above the 0.1\,\%~FAP threshold is located at 5.127 $\pm$ 0.006 d, while no statistically significant signals remain in the final iteration (bottom panel of Figure~\ref{fig:periodogram_b}). Consistent with the decreasing peak power in successive periodograms, the signal amplitude progressively diminishes with each iteration.

Strikingly, all significant peaks can be attributed to known and well-characterized physical processes. The 82.2 $\pm$ 1.6 d signal is associated with the stellar rotation period (Section~\ref{sec:absorption}). The 5.127 $\pm$ 0.006 d period agrees within 1$\sigma$ with the orbital period of Proxima d and may represent an additional manifestation of the proposed star--planet interaction discussed in Section~\ref{sec:d}. The 11.03 $\pm$ 0.03 d signal is consistent, within 1$\sigma$, with a $\sim$2-yr alias of Proxima b’s orbital period. Variations in the primary 44.8-d signal can redistribute power between the 11.03-d peak and its corresponding alias, which is prominent and clearly visible (to the right of the 11.03-d peak) in the second panel of Figure~\ref{fig:periodogram_b}. The proximity of the 11-d signal to Proxima~b’s orbital period together with the absence of this signal in the photospheric features, is suggestive of a planetary-induced modulation of the host star’s strongest emission lines, in agreement with expectations from star--planet magnetic interaction models.  \citet{reville24} suggested that Proxima b lies within the Alfv\'en surface, supporting the enhanced radio emission from Proxima Centauri observed in phase with Proxima b by \citet{pereztorres21}.

An alternative explanation to account for the 11-d signal from Ca\,{\sc ii} H \& K and H$\alpha$ cannot be excluded. In the second panel of Figure~\ref{fig:periodogram_b}, a secondary peak above the 0.1\,\%~FAP threshold is present at $\approx$32.8 d. Including Na\,{\sc i} D1 \& D2 resonance doublet at 5891.583 and 5897.558 \AA~does not modify these findings. The 11-d and 32.8-d signals may share a common origin, as each disappears when the other is removed from the dataset, suggesting that they are not independent periodicities but rather mutually coupled (e.g., harmonics) or aliased components of the same underlying signal. From Mg\,{\sc ii} h$+$k (2800 \AA) emission line observations obtained at a cadence of two measurements per week, \citet{guinan96} reported 20--25\,\%~variability with a period of 31.5 $\pm$ 1.5 d, which they attributed to active plage regions rotating in and out of view on the stellar surface. Accordingly, the 11-d signal detected here may be related to stellar activity occurring on time scales shorter than the full or half stellar rotation period. Alternatively, it might be possible that signatures of a star--planet interaction were already present in those earlier data, although \citet{guinan96} interpreted the periodicity as the stellar rotation period since neither Proxima b nor the true rotation period were known at the time. 

In contrast, the He\,{\sc i} D3 multiplet (5877.24 \AA), despite being a very strong emission line, and He\,{\sc i} $\lambda$6679.996 \AA~behave differently. Their initial joint periodogram peaks near 6.35 d. Subsequent iterations never revealed a signal at $\sim$11.0 d (or its 2-yr alias). Similarly, various Fe\,{\sc i} lines between 5329 and 5458 \AA~did not reveal any signal at 11.0-11.2 d. According to the line formation theory and the expected temperatures of Proxima Centauri's atmosphere, Fe\,{\sc i} emission probes the deepest chromospheric layers near the photosphere; Na\,{\sc i} D lines form just above Fe\,{\sc i} in the lower chromosphere, slightly deeper than Ca\,{\sc ii} H \& K, whose cores trace the lower to mid chromosphere \citep[e.g.,][]{vernazza81, saar97}. Helium emission lines, in contrast, form in the upper chromosphere and corona. If the 11.0-d signal is confirmed to be caused by planet--star magnetic interaction, the energy channeled along magnetic loops from Proxima b to its host star is dissipated in the lower to middle chromosphere, leaving the upper stellar layers apparently unaffected. 

\begin{figure}[h!]
\centering
\includegraphics[width=0.99\hsize]{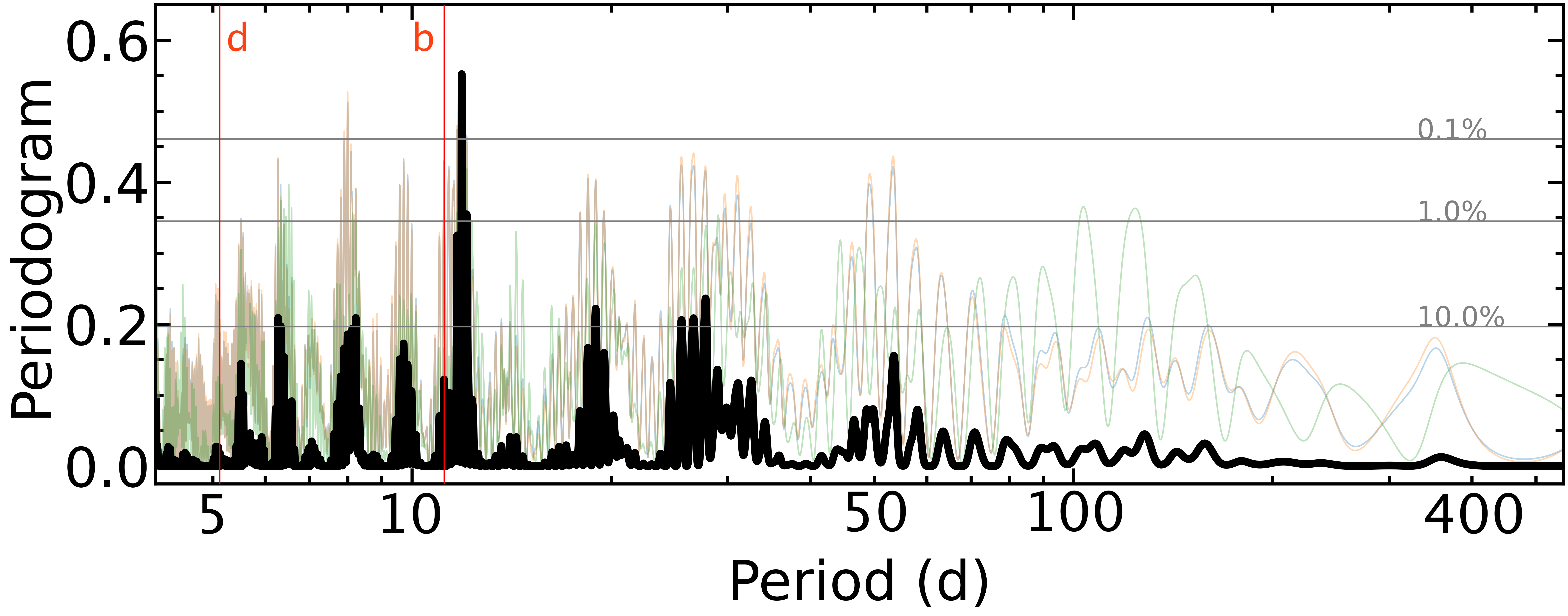}
\caption{Overplotted color periodograms of the Ca\,{\sc ii} H\&K and H$\alpha$ 28 epochs flagged as enhanced iron emission, and the joint periodogram in black. FAP levels are shown by the horizontal gray lines. The vertical red lines denote the orbital periods of Proxima b and d.}
\label{fig:only_emission}
\end{figure}

We caution that the periodogram analysis of the chromospheric lines in this section is highly sensitive to the enhanced-emission (flaring) epochs. When the 28 enhanced iron emission epochs (Section~\ref{sec:d}) are excluded, the 11.03-d and 5.1-d signals disappear, supporting the scenario that these stellar flares may be modulated/driven by Proxima d and b. Crucially, this further supports the planetary origin of the radial velocity (RV) signals \citep{mascareno25}, complementing the evidence from the constant RV amplitude from blue to near-infrared wavelengths, which is expected for planets but not for stellar activity. The joint Ca\,{\sc ii} and H$\alpha$ periodogram of the 28 enhanced iron emission epochs computed between 4.1 d and 550 d has a peak above the 0.1\,\%~FAP at 11.89 $\pm$ 0.06 d (Figure~\ref{fig:only_emission}). This measurement agrees within 1\,$\sigma$ with 11.98 $\pm$ 0.09 d, corresponding to the synodic period between half the chromospheric stellar rotation and $P_1 = 9.454 \pm 0.001$ d, calculated using $P_{\rm syn} = [1/P_1 - 1/P_{\rm rot2}]^{-1}$, where $P_1$ represents the synodic period between planets b and d (see next) and $P_{\rm rot2} = 44.8$ d. The negative sign in the mathematical expression indicates that stellar rotation and the orbital motion of the planets are prograde, that is, the planetary orbital motions and stellar rotation proceed in the same direction. This is consistent with previous dynamical stability studies of the system, which indicate that coplanar, prograde orbits are expected to remain stable \citep[e.g.,][]{livesey24}.

\subsection{Proxima b and d}

\citet{fischer19} proposed another mechanism that could cause temporal variability in magnetic star--planet interactions: the interplay of multiple Alfv\'en wings, known as "wing-wing" interaction. Although Proxima Centauri has a magnetic dipole component \citep{klein21}, the magnetic field likely deviates from a dipolar configuration at large distances from the star, as the stellar wind stretches the field lines outward. If Proxima b and d happened to share the same open field line, the Alfv\'en wings of both planets could interact, with a characteristic period equal to the synodic orbital period of the planets, that is, the period at which their magnetospheres align with the star. For the system, this corresponds to $9.454 \pm 0.001$ d for prograde orbits and $3.5138 \pm 0.0002$ d for prograde and retrograde orbits  (e.g., \citealt{revilla26}). We did not detect strong signals at either of these periods in our study, suggesting that any "wing-wing" interaction in the Proxima Centauri system is either absent during the ESPRESSO observations, below our detection threshold, or not observable due to the global magnetic geometry of the system, which may break any periodicity of the signal. 

If the 11.89-d signal in Figure~\ref{fig:only_emission} instead represented the synodic period between Proxima d and an additional, closer-in planet, both modulating stellar activity in a "wing-wing" interaction scenario, we may predict the orbital period of the hypothetical planet. The expected period would be $\sim$9 d for a prograde orbit and $\sim$3.5 d for a retrograde orbit. This would make the planetary system dynamically packed with strong implications on the dynamical stability. Since no signal is observed in the radial velocity work of \citet{mascareno25}, its mass would be below that of Mars.

\begin{figure*}[h!]
\centering
\includegraphics[width=0.44\hsize]{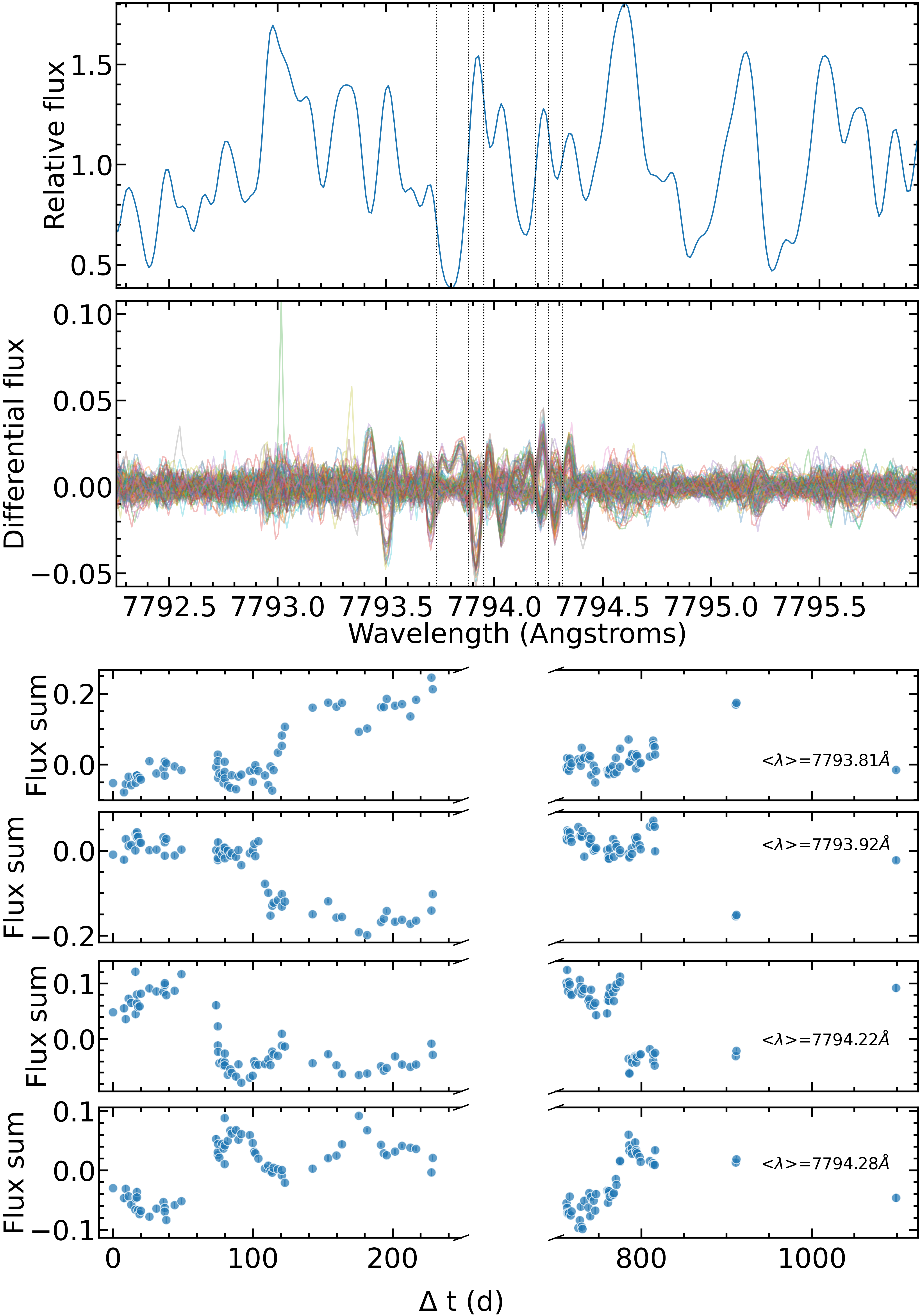} ~
\includegraphics[width=0.44\hsize]{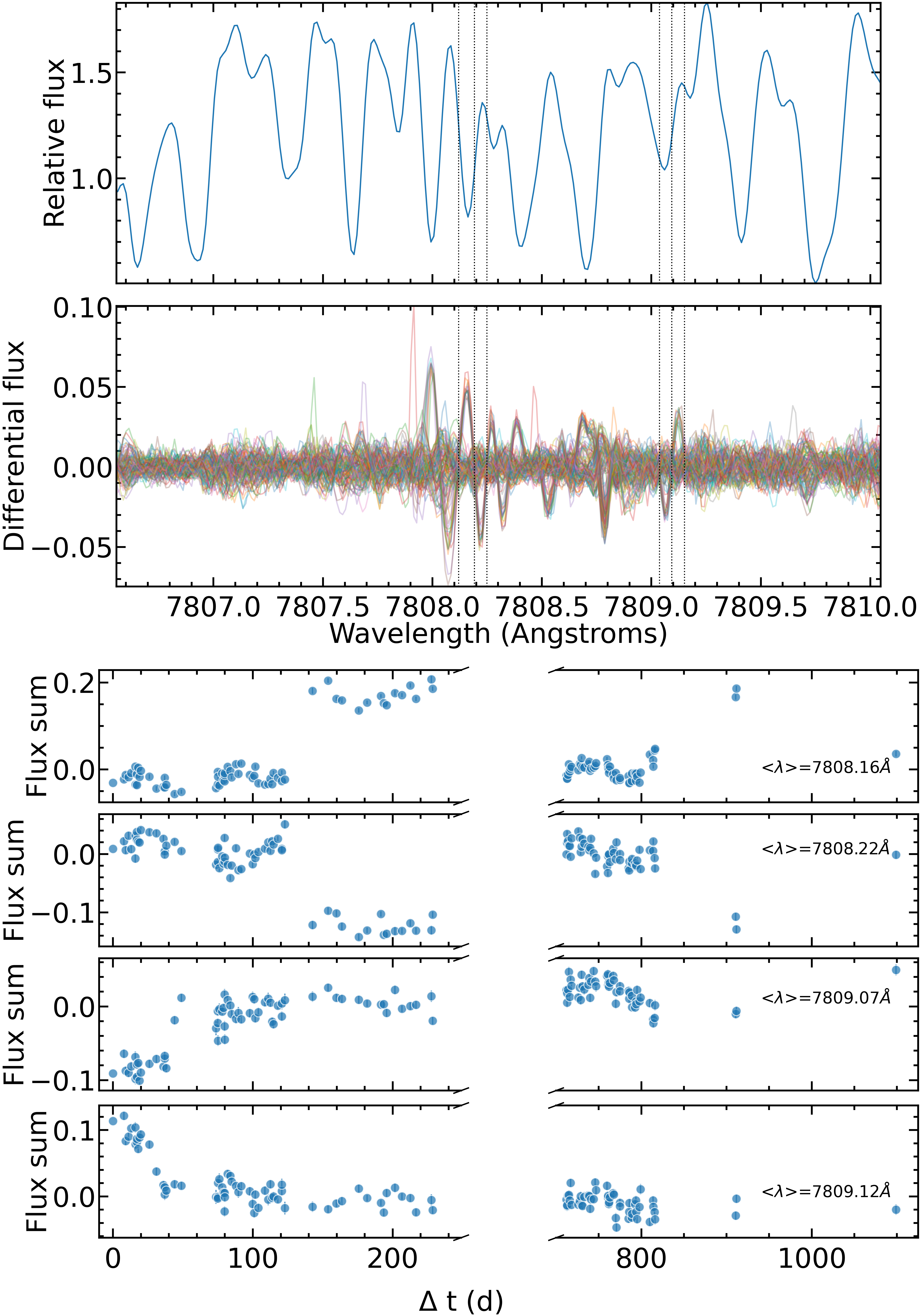}
\caption{A small portion of the combined ESPRESSO (reference) spectrum of Proxima Centauri is shown in the top panels of both columns. Most of the spectral features are due to TiO and VO absorption. The second panels display the 117 individual differential spectra. The vertical black dotted lines mark the wavelength intervals used to compute the summed fluxes displayed in the four bottom panels of each column. Wavelengths are given in vacuum and in the laboratory reference frame.
}
\label{fig:raros}
\end{figure*}

\section{Spectroscopic features with different behavior}

In the differential spectra, several features exhibit variability patterns that notably differ from those of purely photospheric absorption (Figure~\ref{fig:absorption}) and from combined emission-absorption behavior (Figures~\ref{fig:d} and \ref{fig:caii}–\ref{fig:nai}). These anomalous features are predominantly concentrated at red wavelengths, where the measurement precision is highest. Figure~\ref{fig:raros} presents two wavelength regions containing these features  together with their corresponding light curves. While signatures of typical photospheric absorption variability are occasionally visible, the light curves are often dominated by long-term, seasonal trends with duration exceeding one stellar rotation period. For example, the features at 7793.81 and 7808.16 \AA~show low-amplitude variability consistent with photospheric absorption features over the first 110--120 d, followed by an abrupt flux increase of $\approx$20\,\%~lasting at least $\sim$100 d, during which photospheric absorption signatures variability become marginal. A similar cycle is observed in the second half of the time series, albeit with lower observational cadence. Note that the sudden flux increases or decreases in 7793.81, 7793.92, 7794.22, 7794.28, 7808.16, and 7808.22 \AA~depicted in Figure~\ref{fig:raros} occur on time scales shorter than a few tens of days. On the contrary, the feature at 7809.12 \AA~shows a steady flux decrement from the initial observations, reaching a plateau after $\sim$50 d and remaining at this low flux level for the rest of the observing campaigns ($>200$ d). 

Another striking aspect of the light curves in Figure~\ref{fig:raros} is that the seasonal variations are not synchronized across wavelengths. The abrupt flux changes in the 7808.2-\AA\ features occur 10--20 d after those in the 7793.9-\AA\ features, which in turn lag by 30--40 d behind the variations of the 7794.2-\AA\ features. The seasonal change in the 7809.1-\AA\ features occurs earliest. There is no obvious correlation with the 7--8-yr or 18-yr magnetic cycle of Proxima Centauri.

The exact physical mechanism generating these variations is unknown to us. Because only certain molecular transitions show damped-oscillator like variability, the effect possibly arises from magnetic or thermal phenomena rather than changes in molecular abundance. These transitions may have larger effective Land\'e factors, making them more sensitive to time-variable magnetic fields \citep[e.g.,][]{basri92,afram19}. Alternatively, they may form in narrow photospheric layers where propagating waves (e.g., from stellar pulsations) resonate \citep[e.g.,][]{khomenko15}, or correspond to hyperfine components with distinct broadening detectable at high spectral resolution \citep[e.g.,][]{qu22}.

A detailed analysis is beyond the scope of this paper. We report these observational features, which, if confirmed to be linked to stellar magnetic fields, could provide a new way to measure magnetic field strengths in ultra-cool dwarfs from late-M to mid-L spectral types.

\section{Conclusions}

Our analysis of 117 ESPRESSO spectra, originally obtained to confirm the presence of Proxima b orbiting the M5.5 star Proxima Centauri \citep{mascareno20} and subsequently leading to the discovery of the closer-in planet Proxima d \citep{faria22}, showed that the host star was in a flaring state during at least $24.8 \pm 4.7$\,\%~of the total observing time. The study of photospheric features enabled a precise determination of the stellar rotation period and its first harmonic, yielding values ($P_{\rm rot} = 84.9 \pm 0.6$ d) consistent with those reported in the literature. We found no evidence for photospheric variability on time scales comparable to the orbital periods of Proxima b or Proxima d.

In contrast, the chromospheric diagnostics, mainly H$\alpha$, Na\,{\sc i}, and Ca\,{\sc ii} lines, observed in both absorption and emission, showed variability on time scales corresponding to half and full stellar rotation periods as well as to both planetary orbital periods. This behavior suggests that both planets may interact magnetically with their host star and likely remained in a sub-Alfv\'enic regime during (great part of) the ESPRESSO observations. In particular, Proxima d ($P_{\rm orb} = 5.1$ d) appears to induce phase-locked flares of relatively low luminosity ($\sim 10^{24}$--$10^{27}$ erg\,s$^{-1}$), with a statistical significance $\ge 99.8$\,\%. Although this behavior is observed throughout the full ESPRESSO data set, it seems to concentrate during specific epochs. Adopting the helicity-driven reconnection model and associated Poynting flux formalism of \citet{lanza13}, we inferred a polar magnetic field strength of approximately 8--230 G for the Mars-mass Proxima d. This estimate depends on several poorly constrained parameters (e.g., planetary radius and stellar magnetic field strength/geometry) and should be treated as an order-of-magnitude approximation. Although stronger than Earth’s field and weaker than that of giant exoplanets, it remains theoretically plausible for a hot terrestrial planet. Proxima d would thus represent the second known case of a planet inducing stellar flares, following HIP\,67522\,b \citep{ilin25}. The outer planet Proxima b ($P_{\rm orb} = 11.1$ d) does not appear to trigger measurable stellar flares phased with its orbital period, but it may modulate the flare intensity, as suggested by the periodogram analysis of Ca\,{\sc ii} and H$\alpha$ chromospheric lines using both the full set of observing epochs and the subset flagged as enhanced-emission epochs.

We also identified narrow spectral intervals around 7794 \AA~and 7808.5 \AA~that exhibit variability patterns differing from the rest of the photospheric and chromospheric features. These variations might be associated with seasonal changes in Proxima Centauri’s magnetic field, although this interpretation remains unconfirmed.

\begin{acknowledgements}
The authors are grateful to Asensio Ramos, Nikola Vitas, and the rest of the "brown dwarf" team at the IAC for fruitful conversations. The authors also thank the anonymous referee for the insightful review and constructive suggestions that improved the manuscript.
M.R.Z.O$.$, P.A.M.P$.$, V.J.S.B$.$, A.S.M., R.R.L., and J.I.G.H$.$ acknowledge financial support by the Spanish "Ministerio de Ciencia, Innovaci\'on y Universidades" through projects PID2022-137241NB-C42, PID2022-137241NB-C41, and PID2023-149982NB-I00.
P.A.M.P$.$ also acknowledges use of the grant RYC2021-031173-I funded by MCIN/AEI/10.13039/501100011033 and by the European Union NextGenerationEU/PRTR. 
C.J.A.P.M$.$ acknowledges support by Portuguese funds through FCT (Funda\c c\~ao para a Ci\^encia e a Tecnologia) in the framework of the project 2022.04048.PTDC (Phi in the Sky, DOI 10.54499/2022.04048.PTDC). C.J.A.P.M$.$ also acknowledges FCT and POCH/FSE (EC) support through Investigador FCT Contract 2021.01214.CEECIND/CP1658/CT0001 (DOI 10.54499/2021.01214.CEECIND/CP1658/CT0001).
N.C.S$.$ and J.R$.$ is fo-funded by the European Union (ERC, FIERCE, 101052347). Views and opinions expressed are however those of the author(s) only and do not necessarily reflect those of the European Union or the European Research Council. Neither the European Union nor the granting authority can be held responsible for them. 
J.R$.$ is also supported by FCT - Fundação para a Ciência e a Tecnologia through national funds by these grants: UIDB/04434/2020 DOI: 10.54499/UIDB/04434/2020, UIDP/04434/2020 DOI: 10.54499/UIDP/04434/2020, PTDC/FIS-AST/4862/2020, UID/04434/2025.

\end{acknowledgements}

\bibliographystyle{aa}
\bibliography{aa60158-26.bib}

\begin{appendix}

\section{Additional Tables}

\begin{table}[h!]
  \caption{\label{tab:magfield}Proxima d's magnetic field intensity (in G) as a function of stellar magnetic field geometry and intensity, planetary radius, and flare luminosity.}
  \small
\centering
\begin{tabular}{c|c|c|c}
  \hline\hline\noalign{\smallskip}
 $B_*$   & $s=2.0$ & $s=2.5$ & $s=3.0$ \\
 (G) & & &  \\ 
  \hline
  \noalign{\smallskip}
\multicolumn{4}{c}{$P_{\rm SPI} = 6.5\times10^{24}$ erg\,s$^{-1}$} \\
  \hline
200  & \diagbox{10.7}{4.9} & \diagbox{16.0}{7.4} & \diagbox{23.7}{11.0} \\
  \hline
2000 & \diagbox{6.2}{2.8} & \diagbox{9.7}{4.5} & \diagbox{14.7}{6.8} \\
  \hline

\noalign{\smallskip}
  \hline
  \noalign{\smallskip}
\multicolumn{4}{c}{$P_{\rm SPI} = 10^{25}$ erg\,s$^{-1}$} \\
  \hline
200  & \diagbox{13.9}{6.4} & \diagbox{20.8}{9.6} & \diagbox{30.7}{14.3} \\
  \hline
2000 & \diagbox{8.1}{3.7} & \diagbox{12.7}{5.8} & \diagbox{19.1}{8.8} \\
  \hline
  
\noalign{\smallskip}
  \hline
  \noalign{\smallskip}
\multicolumn{4}{c}{$P_{\rm SPI} = 4\times10^{26}$ erg\,s$^{-1}$} \\
  \hline
200  & \diagbox{130.1}{60.3} & \diagbox{192.2}{89.4} & \diagbox{282.0}{131.5} \\
  \hline
2000 & \diagbox{79.2}{36.3} & \diagbox{119.3}{55.3} & \diagbox{176.7}{82.2} \\
  \hline
\noalign{\smallskip}
\end{tabular}
\tablefoot{In each cell, the top-right value is for Earth-size and the bottom-left for Mars-size radius.}

\end{table}

\section{Additional Figures}

\begin{figure}[h!]
\centering
\includegraphics[width=0.99\hsize]{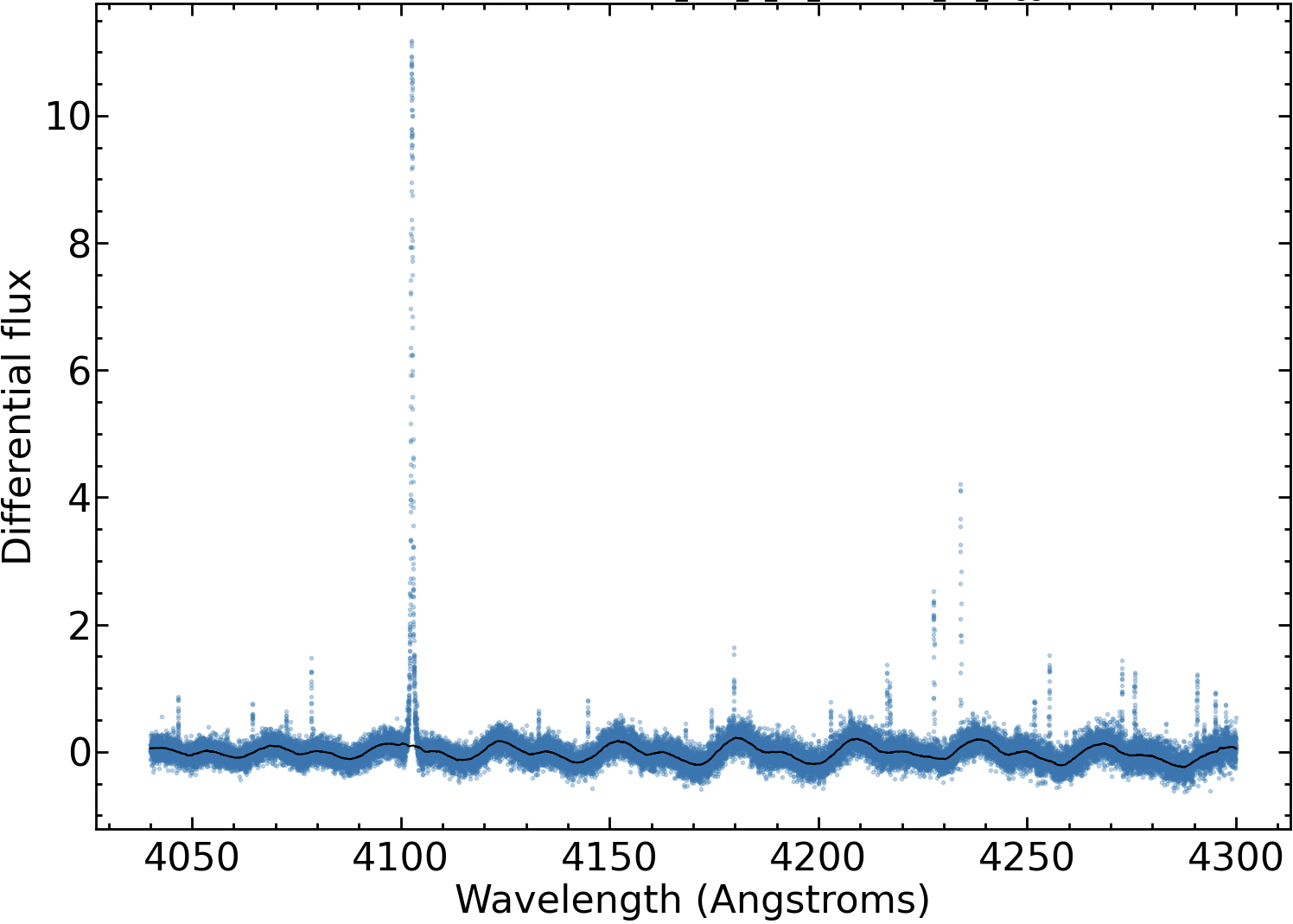} 
\caption{An example of the long-period "wiggles" correction applied to the differential spectra (blue) is shown. The black line represents the function used to remove the "wiggle" interference pattern without affecting the narrow-band absorption and emission features. The strongest emission is due to H$_\delta$.
}

\label{fig:wiggles}
\end{figure}

\begin{figure}[h!]
\centering
\includegraphics[width=0.99\hsize]{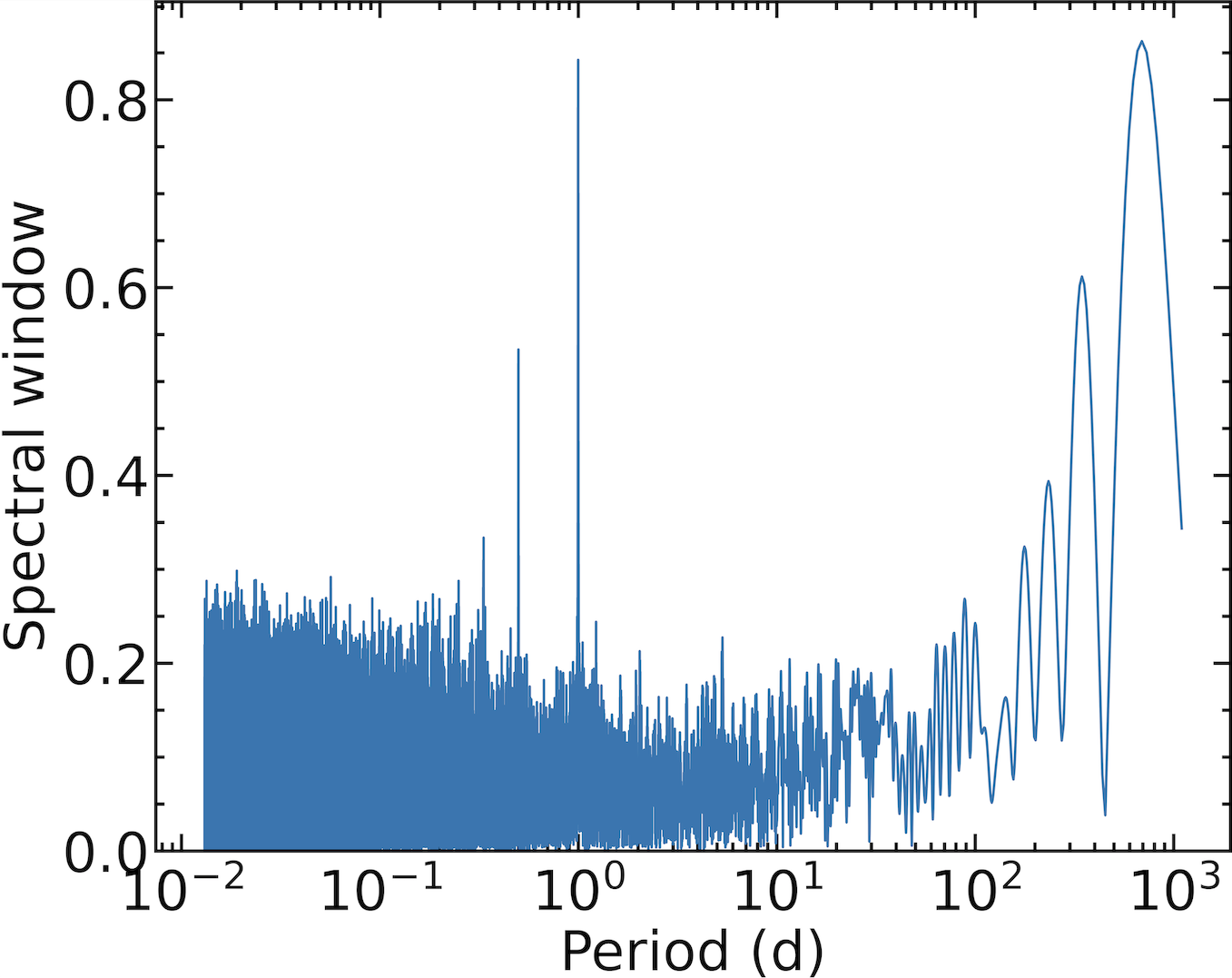} 
\caption{Spectral window of the ESPRESSO data based on all 117 epochs.
}

\label{fig:spectral_window}
\end{figure}

\begin{figure}[h!]
\centering
\includegraphics[width=0.99\hsize]{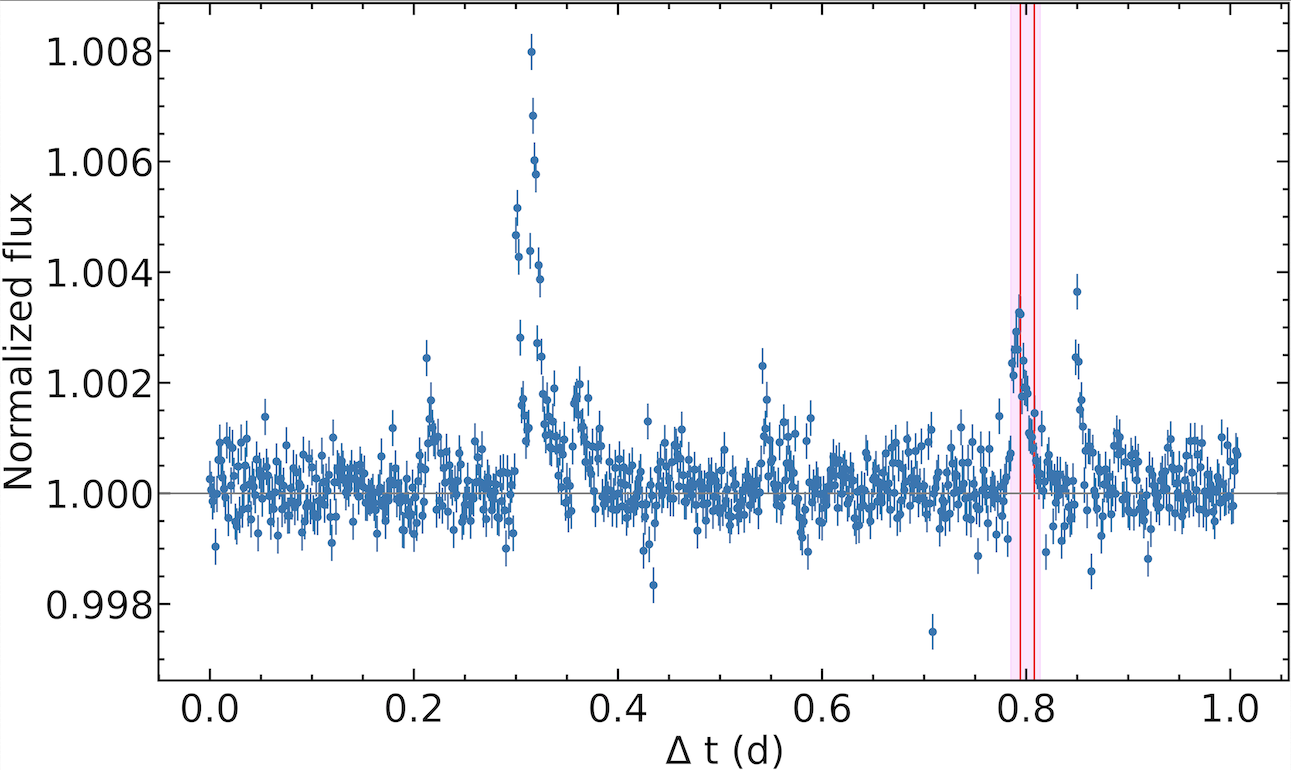} 
\caption{Normalized one-day segment of the {\sl TESS} light curve (Sector 11) of Proxima Centauri. The two ESPRESSO epochs corresponding to enhanced emission and coincident with a {\sl  TESS} flare are marked by red vertical lines (2019 April 26). The duration of the event is highlighted in magenta.
}
\label{fig:tess}
\end{figure}

\begin{figure}[h!]
\centering
\includegraphics[width=0.99\hsize]{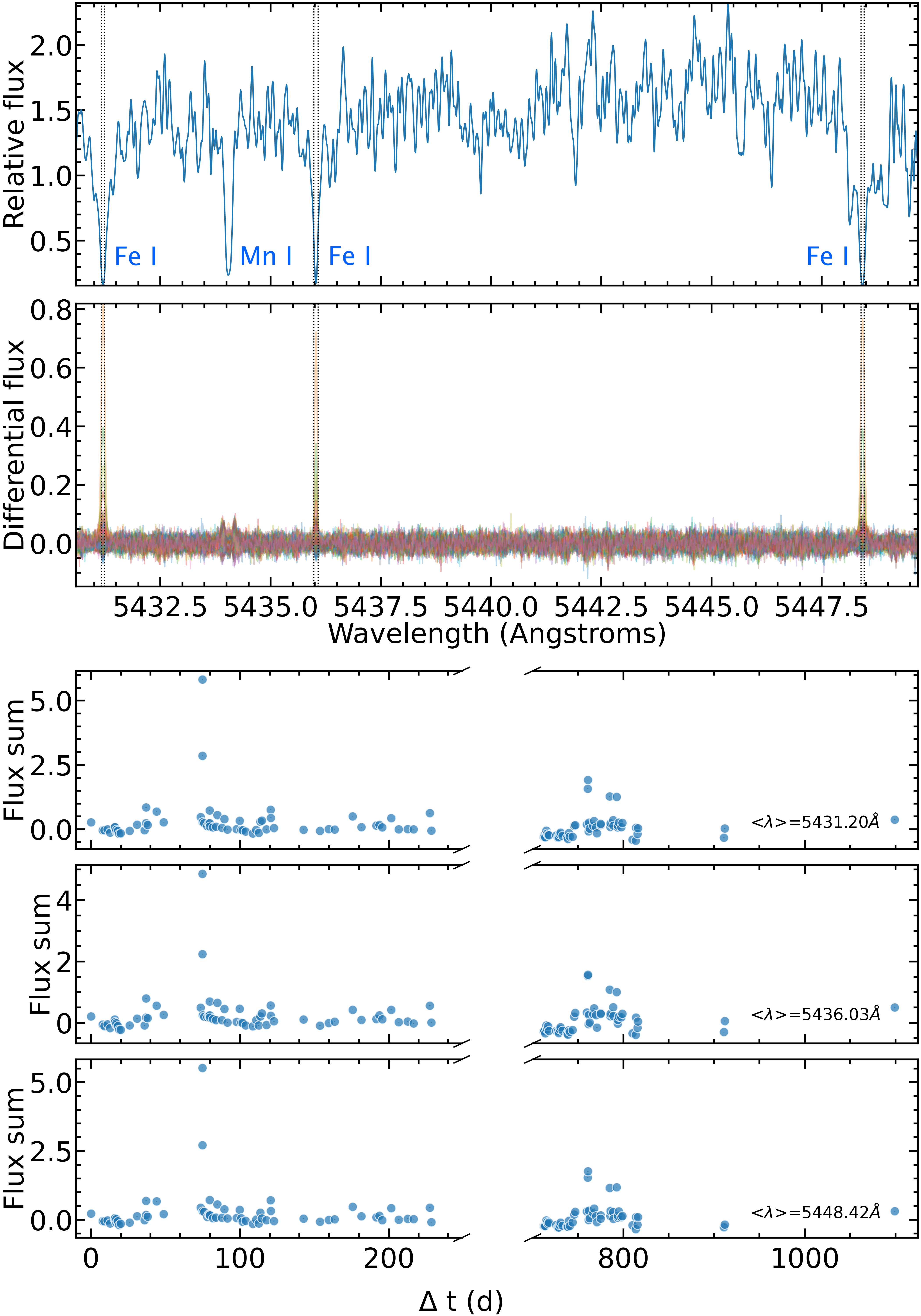}
\caption{A small portion of the combined ESPRESSO (reference) spectrum of Proxima Centauri is shown in the first panel. A few atomic lines are identified. The second panel displays the 117 individual differential spectra, where iron lines appear either in absorption or emission. The vertical black dotted lines indicate the size of the features used to compute the summed fluxes shown in the three bottom panels. Wavelengths are in vacuum and in the laboratory reference frame.
}
\label{fig:iron}
\end{figure}

\begin{figure}[h!]
\centering
\includegraphics[width=0.99\hsize]{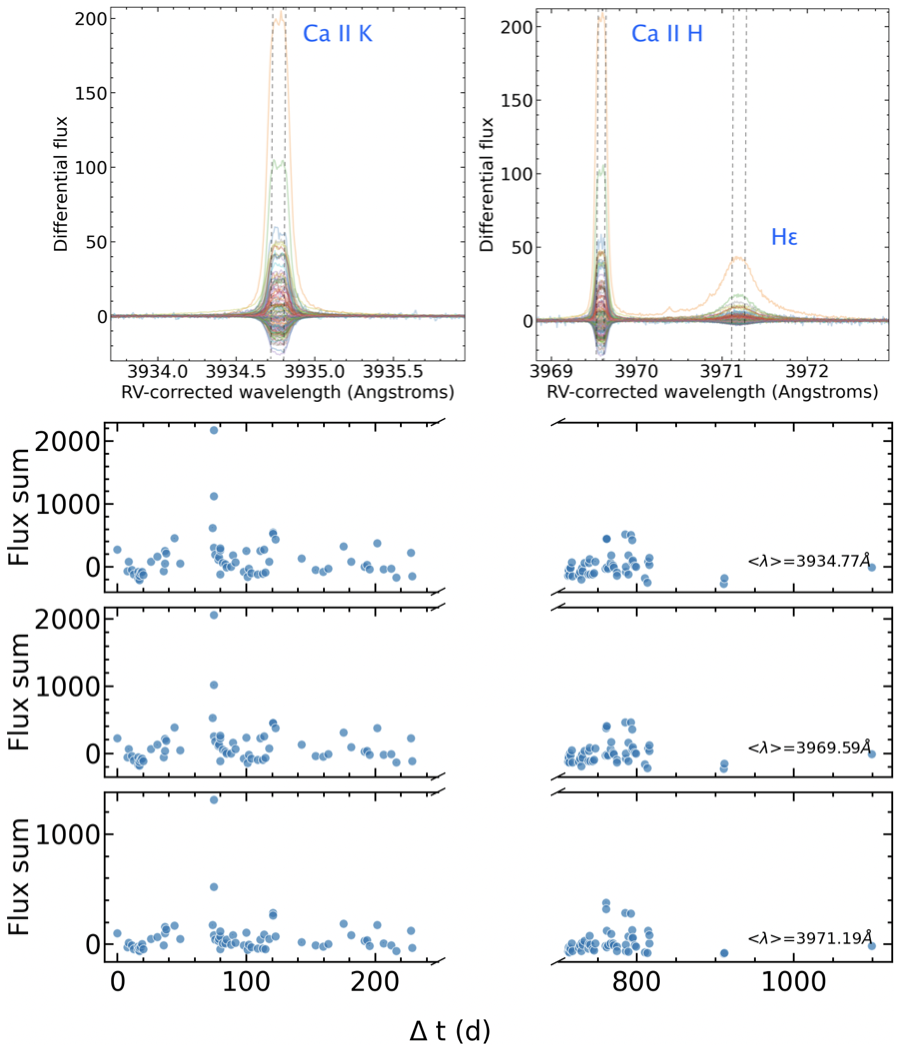}
\caption{A small portion of the combined ESPRESSO (reference) spectrum of Proxima Centauri is shown in the first panel. The strong emission lines are due to Ca\,{\sc ii} H \& K and H$\epsilon$. The second panel displays the 117 individual differential spectra. The vertical black dotted lines indicate the size of the features used to compute the summed fluxes shown in the three bottom panels. Wavelengths are in vacuum and in the laboratory reference frame.
}
\label{fig:caii}
\end{figure}

\begin{figure}[h!]
\centering
\includegraphics[width=0.99\hsize]{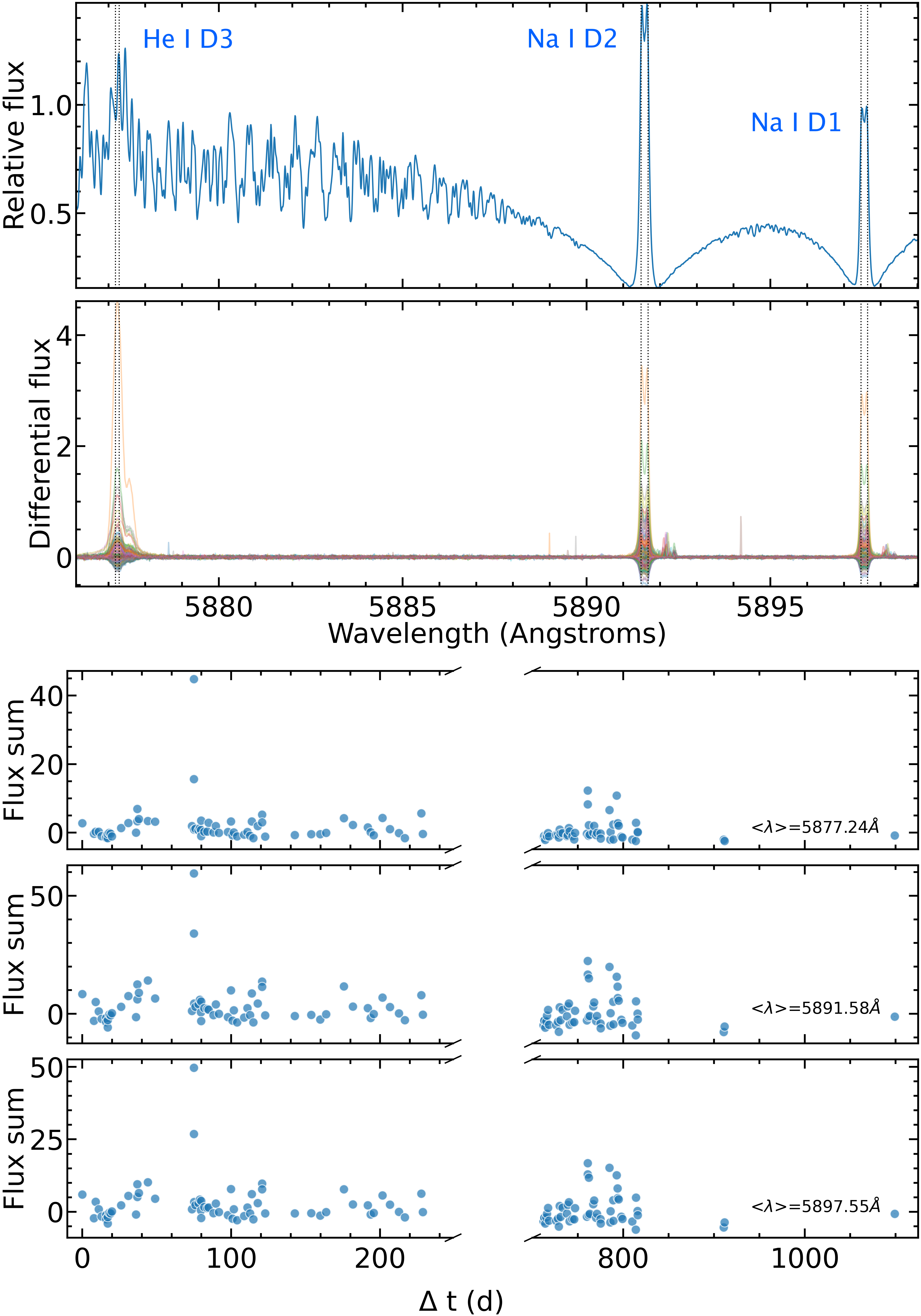}
\caption{A small portion of the combined ESPRESSO (reference) spectrum of Proxima Centauri is shown in the first panel. A few atomic lines are identified. The second panel displays the 117 individual differential spectra, where He\,{\sc i} and Na\,{\sc i} lines appear either in absorption or emission. The vertical black dotted lines indicate the size of the features used to compute the summed fluxes shown in the three bottom panels, where the most intense flare has been removed for clarity. Wavelengths are in vacuum and in the laboratory reference frame. The small "emission" features to the red of the Na\,{\sc i} signatures are telluric residuals that do not affect the analysis.
}
\label{fig:nai}
\end{figure}

\begin{figure}[h!]
\centering
\includegraphics[width=0.99\hsize]{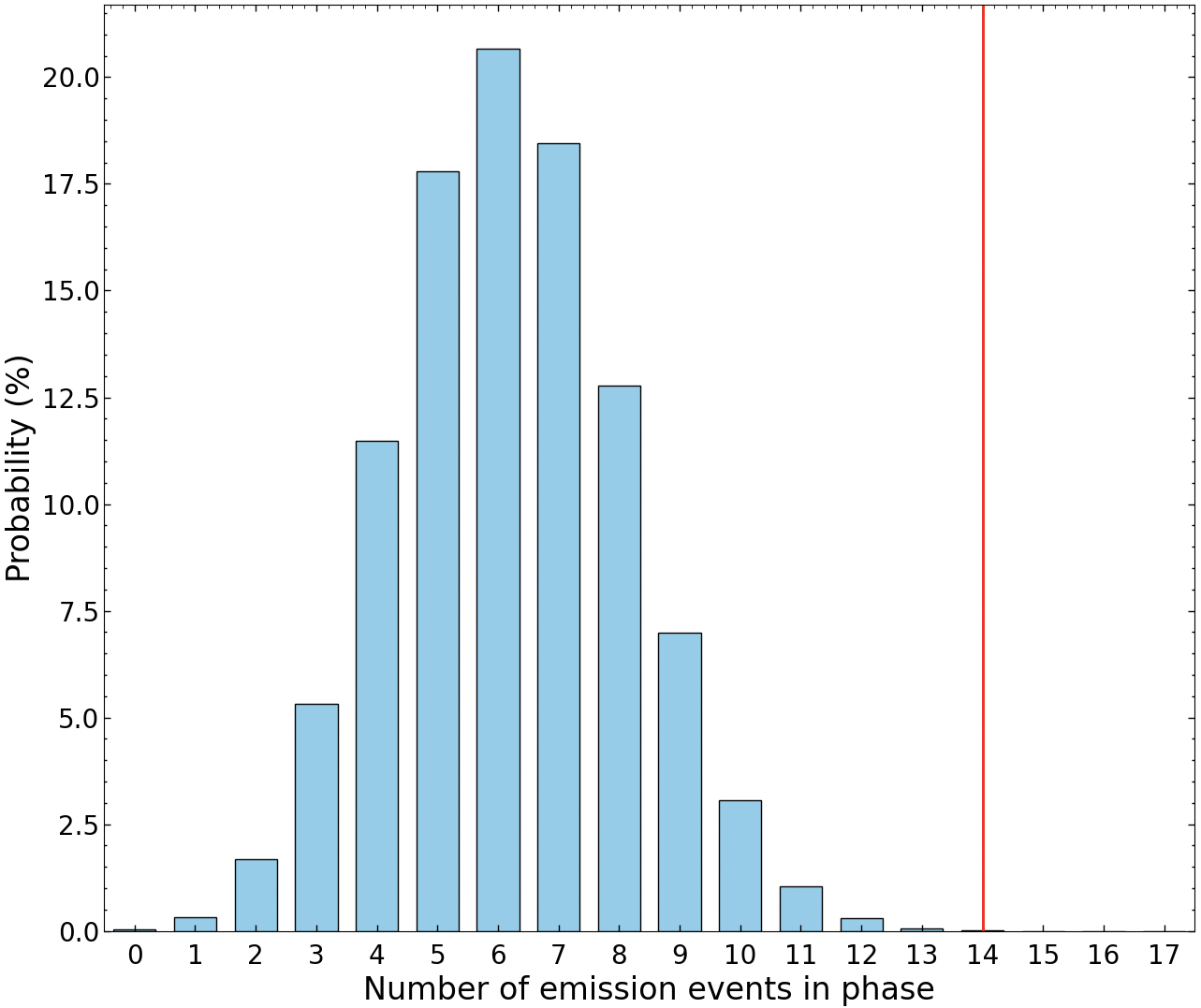}
\caption{Probability distribution of finding enhanced-emission events phased with Proxima d (adopted $t_o = 2458604.61011$ BJD) using the exact arrangement of ESPRESSO observing times, a random redistribution of 28 emission flags, and 10$^6$ iterations. The vertical red line stands for the 14 events identified in the real observations.}
\label{fig:mc_permutation}
\end{figure}

\begin{figure*}[h!]
\centering
\includegraphics[width=0.99\hsize]{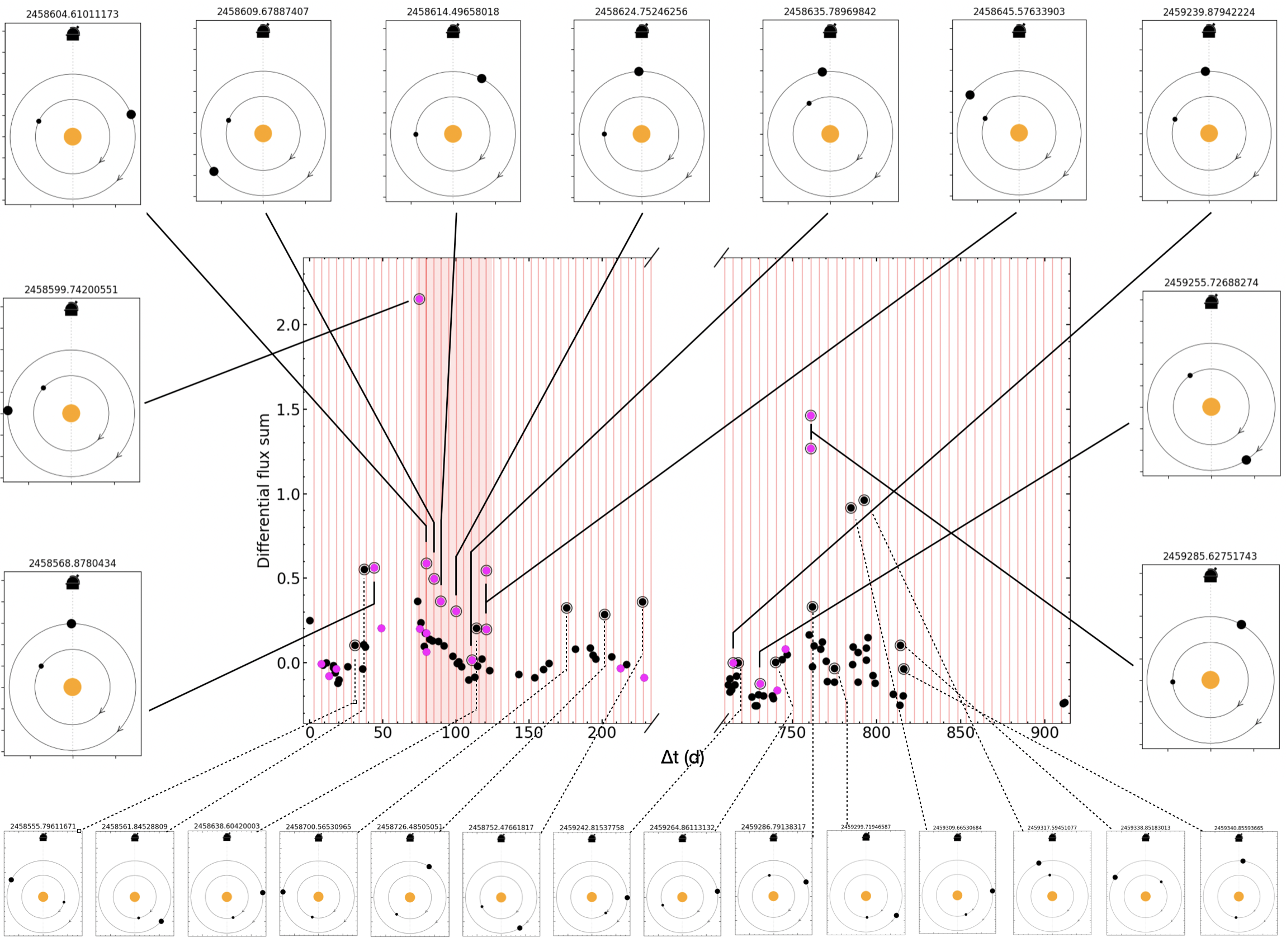}
\caption{Pole-on views of the orbital positions of Proxima d (small black dot) and b (large black dot) during each enhanced-emission event. The telescope symbol and vertical dotted line indicate the time of inferior conjunction and the observer’s direction. Orbital motion is clockwise. Each configuration is labeled by its BJD. Proxima Centauri is represented by the central orange dot. The central panel is identical to Figure~\ref{fig:d}.
}
\label{fig:cartoon}
\end{figure*}

\end{appendix}

\end{document}